\documentclass[useAMS,usenatbib,letterpaper]{mn2e}
 \usepackage[totalwidth=500pt,totalheight=630pt,centering,layoutvoffset=-0.3cm]{geometry}
\usepackage{amssymb}
\usepackage{aas_macros}
\usepackage{natbib}
\usepackage{times}
\usepackage{graphicx}

\newcommand{\Msun}                    {\,{\rm M}_\odot}

\newcommand{\hkpc}                     {\,h^{-1}\,{\rm kpc}}
\newcommand{\hMpc}                    {\,h^{-1}\,{\rm Mpc}}
\newcommand{\hMsun}                  {\,h^{-1}\,{\rm M}_\odot}
\newcommand{\kms}                      {\,\,{\rm km}\,\,{\rm s}^{-1}}

\newcommand{\gimic}        {\textsc{gimic}}
\newcommand{\owls}          {\textsc{owls}}
\newcommand{\gadget}      {\textsc{Gadget3}}

\newcommand{\newl} 	{\mathrm{log_{10}}}

\def\lsim{\mathrel{\lower0.6ex\hbox{$\buildrel {\textstyle <}
 \over {\scriptstyle \sim}$}}}

\title[Impact of galaxy formation on halo properties]{The impact of galaxy formation on the total mass, mass profile and abundance of haloes}%
 \author[M. Velliscig et al.]{Marco Velliscig$^{1}$\thanks{E-mail: velliscig@strw.leidenuniv.nl}, Marcel P. van Daalen$^{1,2}$, Joop Schaye$^{1}$, Ian G. McCarthy$^{3}$, \newauthor
~Marcello Cacciato$^{1}$, Amandine~M.~C.~Le Brun$^{3}$ and Claudio Dalla Vecchia$^{4,5}$ \\\\
  $^{1}$Leiden Observatory, Leiden University, P.O. Box 9513, 2300 RA Leiden, The Netherlands \\
  $^{2}$Max Planck Institute for Astrophysics, Karl-Schwarzschild Strasse 1, 85741 Garching, Germany\\
  $^{3}$Astrophysics Research Institute, Liverpool John Moores University, 146 Brownlow Hill, Liverpool L3 5RF\\
  $^{4}$Instituto de Astrof\'isica de Canarias, C/ V\'ia L\'actea s/n, 38205 La Laguna, Tenerife, Spain\\
  $^{5}$Departamento de Astrof\'sica, Universidad de La Laguna, Av. del Astrof\'isico Franciso S\'anchez s/n, 38205 La Laguna, Tenerife, Spain
}

\begin{document}

\date{\today}
\pagerange{\pageref{firstpage}--\pageref{lastpage}} \pubyear{2014}

\maketitle

\label{firstpage}
\begin{abstract}
We use cosmological hydrodynamical simulations to investigate how the inclusion of physical processes relevant to galaxy formation (star formation, metal-line cooling, stellar winds, supernovae and feedback from Active Galactic Nuclei, AGN) change the properties of haloes, over four orders of magnitude in mass. We find that gas expulsion and the associated dark matter (DM) expansion induced by supernova-driven winds are important for haloes with masses $M_{200} \la 10^{13}\Msun$, lowering their masses by up to $20\%$ relative to a DM-only model. AGN feedback, which is required to prevent overcooling, has a significant impact on halo masses all the way up to cluster scales ($M_{200} \sim 10^{15}\Msun$). Baryon physics changes the total mass profiles of haloes out to several times the virial radius, a modification that cannot be captured by a change in the halo concentration. The decrease in the total halo mass causes a decrease in the halo mass function of about $20\%$. This effect can have important consequences for the abundance matching technique as well as for most semi-analytic models of galaxy formation. We provide analytic fitting formulae, derived from simulations that reproduce the observed baryon fractions, to correct halo masses and mass functions from DM-only simulations. The effect of baryon physics (AGN feedback in particular) on cluster number counts is about as large as changing the cosmology from WMAP7 to Planck, even when a moderately high mass limit of $ M_{500}\approx10^{14}\Msun$ is adopted. Thus, for precision cosmology the effects of baryons must be accounted for.
\end{abstract}


\section{Introduction}
\label{sec:introduction}
N-body simulations using only gravitationally interacting dark matter (DM) particles have played an important role in the development of astrophysical cosmology \citep[e.g.][]{Frenk12}. 
DM simulations have for example been used to predict the large-scale distribution of matter, the halo mass function, and the density profiles of haloes. The results from such simulations form the basis for halo-based models and abundance matching techniques \citep[e.g.][]{Seljak00,Cooray02}. DM simulations are also used as the starting point for semi-analytic models that associate
galaxies to DM haloes in post-processing, and then follow the evolution of these galaxies according to different prescriptions that are calibrated such that the model reproduces a limited set of observables \citep[e.g.][]{Baugh06}.

Neglecting the hydrodynamics and the feedback processes that affect the gas component also allows one to perform simulations with a dynamic range that would otherwise not be achievable due to the higher computational cost associated with the inclusion of baryon physics. 
Explicitly accounting for baryons in simulations means computing hydrodynamic forces and including processes like radiative cooling, photo-heating, star formation, metal enrichment, and also energetic feedback processes such as winds driven by supernovae and AGN that are able to generate outflows and eject baryons from (the inner parts of) DM haloes. 

For many purposes it is a reasonable approximation to assume that the effect of baryon physics on the matter distribution is small, such as in the intergalactic medium (e.g., \citealt{Theuns02,Viel12}) and on the outskirts of galaxy clusters where the gas has a long cooling time and approximately traces the dark matter (e.g., \citealt{LeBrun14}).  However, it is clear that on small scales and in lower mass haloes, where cooling can allow the gas to condense to high densities, baryonic processes such as galactic winds can have important effects on the matter distribution.  Moreover, the distribution of the DM will itself adjust to the resulting change in the gravitational potential. Indeed, it appears that the observed rotation curves of dwarf galaxies cannot be reproduced by simulations that assume the standard cold dark matter paradigm unless they include the effect of outflows \citep[e.g.][]{Governato10}. 

In recent years hydrodynamical simulations have for example been used to quantify the effect of baryons on the DM halo density profiles \citep[e.g.][]{Gnedin04,Gustafsson06,Duffy10,Tissera10}, spins \citep[e.g.][]{Bett10,Bryan12}, shapes \citep[e.g.][]{Abadi10,Kazantzidis04,Read08,Bryan11,Bryan12}, and substructure \citep[e.g.][]{Dolag09,Romano09} of dark haloes, as well as on the matter power spectrum \citep[e.g.][]{Jing06,Rudd08,Guillet10,VanDaalen11,Casarini11} and the clustering of subhaloes \citep{VanDaalen13}. Because the physics of galaxy formation is uncertain, it is important to vary the parameters of the model. In particular, it has recently become clear that the efficient feedback that is required to reproduce observations, and which is thought to be driven by star formation and by AGN at low and high halo masses, respectively, leads to results that differ qualitatively from the predictions of earlier simulations that suffered from overcooling. For example, efficient feedback reduces, or even reverses, adiabatic contraction in the inner parts of massive haloes \citep[e.g.][]{Duffy10,Mead10,Teyssier11,Killedar12,Martizzi13a}. The ejection of baryons by outflows reduces the matter power spectrum on remarkably large scales \citep{VanDaalen11} with dramatic consequences for future cosmological weak lensing studies \citep{Semboloni11,Semboloni12,Zentner13}.

One of the most important quantities characterising the distribution of matter is the halo mass function (HMF hereafter), i.e.\ the number density of haloes as a function of their mass. The evolution of the HMF, in particular its massive end, is for example a powerful tool for constraining cosmological parameters such as the dark energy equation of state, using future large surveys such as eROSITA \citep{eRosita12}, XMM-XXL \citep{Pierre11}, and XCS \citep{Mehrtens12} in X-ray; Planck\footnote{http://www.rssd.esa.int/index.php?project=Planck} using the Sunyaev-Zel'dovich effect; and DES\footnote{http://www.darkenergysurvey.org/}, Euclid \citep{Euclid11} and LSST\footnote{http://www.lsst.org/lsst/} using weak lensing. For a given cosmology, the HMF is usually predicted using DM-only simulations
\citep[e.g.][]{Jenkins01,Reed03,Warren06,Lukic07,Tinker08}. To exploit the capacity of upcoming surveys, the theoretical HMF needs to be calibrated at the per cent level \citep{Wu10}. Therefore, even if the impact of baryon physics is only of the order of a per cent or more, it should be taken into account when computing the theoretical HMF.

The effect of baryon physics on the HMF  has recently been studied using hydrodynamical simulations by, among others, \citet{Sawala12} and \citet{Khandai14} at low halo masses and by \citet{Stanek09}, \citet{Cui12}, \citet{Martizzi13b} and \citet{Cusworth13} at high masses.  Although the studies differ in their detailed findings (and sometimes on the sign of the effect), there is nevertheless a growing consensus that baryon physics will significantly affect the HMF.

Here we will use the suite of cosmological simulations from the OverWhelmingly Large Simulations project (OWLS; \citealt{Schaye10}) to study the effect of galaxy formation on the mass function and internal structure of haloes more massive than $M^{\mathrm{crit}}_{200} = 10^{11.5} \hMsun$. For this purpose we will also make use of new larger volume, lower resolution versions of a subset of the OWLS models (an extension of OWLS, called cosmo-OWLS; see \citealt{LeBrun14}). OWLS is well-suited to study baryonic effects as it consists of a wide range of models that were run from identical initial conditions, but employing a wide variety of recipes for the uncertain baryonic processes. OWLS also includes a DM-only model as well as a model with AGN feedback that reproduces both optical and X-ray observations of groups and clusters of galaxies \citep{McCarthy10,LeBrun14}. These last two models are therefore particularly well-suited to our needs and we will employ them to provide fitting formulas that can be used to correct the HMFs predicted by DM-only models for the effect of baryons.

This paper is organised as follows.  In Sec.~\ref{sec:OWLS} we describe our simulations and explain the methods used to find haloes and to match them between different simulations. In Sec.~\ref{sec:result} we show how baryon physics, such as radiative cooling and feedback from supernovae and AGN alter the masses of haloes.  In Sec.~\ref{Sec:fit_formula} we provide analytic fitting formulae to correct the masses of haloes in the DM-only simulation for the effect of baryon physics. In Sec.~\ref{sec:hmf} we show the impact of baryon physics on the halo mass function and discuss the implications for cluster number count studies.
In Sec.~\ref{sec:discussion} we compare our findings to those of previous studies.  Finally, in Sec.~\ref{sec:conclusions} we summarize and conclude.

\section{Simulations}\label{sec:OWLS}
\begin{table*} 
\begin{center}

\begin{tabular}{lrrcccl}
\hline
Simulation  & L & $N_{\rm particle}$& Cosmology & $m_{\mathrm{b}}  $ & $m_{\mathrm{dm}}$ & Description\\ 
 & $(\hMpc)$ & &  & $ (\hMsun) $ & $(\hMsun)$ & \\
\hline 
\emph{DMONLY} & 100  & $512^3$ & WMAP3 & -- &$ 4.9 \times 10^8$& Only gravitationally interacting particles\\
\emph{NOSN\_NOZCOOL} &100  & $2 \times 512^3$ & WMAP3  & $ 8.7 \times 10^7$  &$ 4.1 \times 10^8$ & No SN feedback, primordial cooling\\
\emph{NOZCOOL} & 100  & $2 \times 512^3$ & WMAP3  & $ 8.7 \times 10^7$  &$ 4.1 \times 10^8$ & Cooling assumes primordial abundances\\
\emph{REF} &  100  & $2 \times 512^3$ & WMAP3 & $ 8.7 \times 10^7$  &$ 4.1 \times 10^8$ & SN feedback, metal line cooling, no AGN\\
\emph{WDENS} & 100  & $2 \times 512^3$ & WMAP3  & $ 8.7 \times 10^7$  &$ 4.1 \times 10^8$ & Wind mass loading and velocity depend on $\rho_{\mathrm{gas}}$\\
\emph{AGN 8.0} & 100  & $2 \times 512^3$ & WMAP3 &$ 8.7 \times 10^7$  &$ 4.1 \times 10^8$ & Includes AGN\\
\hline
\emph{DMONLY L050N512} & 50  & $512^3$ & WMAP3 & -- &$ 6.2 \times 10^7$ & High-res version of DMONLY, smaller box \\
\emph{REF L050N512} & 50  & $2 \times 512^3$ & WMAP3 &$ 1.1 \times 10^7$  &$ 5.1 \times 10^7$ & High res version of REF, smaller box \\
\hline
\emph{DMONLY W7} & 100  & $512^3$ & WMAP7 &  -- &$ 5.6 \times 10^8$ &Different cosmology w.r.t. DMONLY\\
\emph{REF W7} & 100  & $2 \times 512^3$ & WMAP7 & $ 9.4 \times 10^7$  &$ 4.7 \times 10^8$ &Different cosmology w.r.t. REF\\
\emph{AGN 8.0 W7} & 100  & $2 \times 512^3$ & WMAP7 & $ 9.4 \times 10^7$  &$ 4.7 \times 10^8$ &Different cosmology w.r.t. AGN 8.0\\
\emph{AGN 8.5 W7} & 100  & $2 \times 512^3$ & WMAP7 & $ 9.4 \times 10^7$  &$ 4.7 \times 10^8$ &Different heating temperature w.r.t. AGN 8.0 W7\\
\hline
\emph{DMONLY L400 W7} & 400  & $1024^3$ & WMAP7 & --  &$ 4.5 \times 10^9$ & Larger box, lower res w.r.t. DMONLY W7\\
\emph{REF L400 W7} & 400  & $2 \times 1024^3$ & WMAP7 & $ 7.5 \times 10^8$  &$ 3.7 \times 10^9$ & Larger box, lower res w.r.t. REF W7\\
\emph{AGN 8.0 L400 W7} & 400  & $2 \times 1024^3$ & WMAP7 & $ 7.5 \times 10^8$  &$ 3.7 \times 10^9$ & Larger box, lower res w.r.t. AGN 8.0 W7\\
\emph{AGN 8.5 L400 W7} & 400  & $2 \times 1024^3$ & WMAP7 & $ 7.5 \times 10^8$  &$ 3.7 \times 10^9$ & Larger box, lower res w.r.t. AGN 8.5 W7\\
\hline
\end{tabular}
\caption{List of the simulations used in this work. Most simulations use a box of $100 \hMpc$ , with $2 \times  512^3$ particles. We carry out resolution tests using simulations with 8 times higher and lower mass resolution. We also use simulations with a different cosmology, WMAP7 instead of WMAP3, to see if our analysis is cosmology dependent. Finally, we take advantage of  $400 \hMpc$ , $2 \times  1024^3$ version of the OWLS models to extend our analysis to higher halo masses.} 
\label{tbl:sims} 
\end{center}
\end{table*}

The analysis carried out in this paper is based on simulations that are part of the OverWhelmingly Large Simulations project (\owls \, \citealp{Schaye10}) which includes over 50 large, cosmological, hydrodynamical simulations run with a modified version of the smoothed particle hydrodynamics (SPH) code \gadget \, (last described in \citealp{Springel05_gadget}). The aim of the \owls \,project is to explore the sensitivity of the theoretical predictions to both resolvable and `subgrid' physics thought to be important for galaxy formation (such as supernova (SN) feedback, stellar mass loss, radiative cooling processes and AGN feedback) in fully self-consistent cosmological hydrodynamical simulations.  In this section we will give a brief description of the simulations used in this paper and the physical processes implemented in each of them.

The simulations used in this work were run with either a WMAP3 cosmology \citep{wmap3} \{$\Omega_{m}$, $\Omega_{b}$, $\Omega_{\Lambda}$, $\sigma_{8}$, $n_{s}$, $h$\} = \{0.238,0.0418,0.762,0.74,0.951,0.73\} or a WMAP7 cosmology \citep{wmap7} \{$\Omega_{m}$, $\Omega_{b}$, $\Omega_{\Lambda}$, $\sigma_{8}$, $n_{s}$, $h$\} = \{0.272, 0.0455, 0.728, 0.81, 0.967, 0.704\}.
Most simulations used in this work were run in periodic boxes of $100\hMpc$ ($400\hMpc$) comoving, and each of the runs uses $512^3$ ($1024^3$) dark matter and equally many baryonic particles (representing collisionless star or collisional gas particles). The particle masses in the $2\times512^3$ particle 100~$\hMpc$ WMAP3 ($1024^3$ particle $400\hMpc$ WMAP7) simulations are $4.06\times 10^8\,\hMsun$ ($3.75\times 10^9\,\hMsun$) for dark matter and  $8.66\times10^7\,\hMsun$ ($7.53\times 10^8\,\hMsun$) for baryons.  Note, however, that baryonic particle masses change during the course of the simulation due to mass transfer from star to gas particles.

Comoving gravitational softenings were set to $1/25$ of the initial mean inter-particle spacing but were limited to a maximum physical scale of $2\hkpc$ ($4\hkpc$) for the $100\hMpc$ ($400\hMpc$) simulations. The switch from a fixed comoving to a fixed proper softening happens at $z=2.91$ in all simulations. We used $N_{\rm ngb} = 48$ neighbours for the SPH interpolation.   

The physical models considered here are (following the naming convention of \citealp{Schaye10}):

\begin{itemize}
\item DMONLY: a dark matter only simulation, intended to simulate a set of particles that interact only gravitationally. Such simulations are commonly used to compute the HMF that forms the input of semi-analytic models and abundance matching studies. We use this simulation as a base and evaluate differences with respect to this model when baryon physics is added. Recall that a particle in this simulation becomes two particles in a baryonic simulation: namely a DM particle of mass $  \frac{\Omega_{\rm m}-\Omega_{\rm b}}{\Omega_{\rm m}}\times m^{\mathrm{dmonly}}$ and a gas particle of mass $  \frac{\Omega_{\rm b}}{\Omega_{\rm m}}\times m^{\mathrm{dmonly}}$.
\item REF: this is the reference model for the OWLS suite, but is not intended to be the `best' model.  This model includes most of the mechanisms that
  are thought to be important for the star formation history (see \citealp{Schaye10} for a detailed discussion), but not AGN feedback. The implementation of radiative cooling, star formation, supernova driven winds, and stellar evolution and mass loss have been described in \citet{Wiersma09a}, \citet{Schaye08}, \citet{DallaVecchia08}, and \citet{Wiersma09b}, respectively.  This simulation represents a standard scenario assumed in cosmological hydrodynamic simulations. The SN feedback is kinetic and is performed by kicking the particles stochastically in random directions. The parameters that regulate the feedback process are the mass loading $\eta=2$, which represents the
average number of particles kicked per star particle in the case
of equal mass particles, and the initial wind velocity $v_{\rm{w}}=600\kms$. In this simulation the wind parameters are kept fixed and correspond to an injection of energy that is $40\%$ of the available energy of the SN explosion.
\item AGN 8.0: this model is identical to REF with the exception that it also includes a prescription for black hole (BH) growth and AGN feedback, following \citet{Booth09}. In this approach, which is a modified version of the one introduced by \citet{Springel05_agn}, the accretion of gas on to the BH follows the Bondi-Hoyle accretion formula only if the gas is expected to be warm (i.e. $\gtrsim10^4 K$). However, if the pressure is sufficiently high that a cold interstellar phase is expected to form, but which is unresolved by our simulations, then the accretion is regulated by a parameter that depends on the density of the gas, multiplied by the Bondi-Hoyle accretion rate. A certain fraction of the rest mass energy of the accreted gas, $\epsilon$, is stored until it is able to heat up a number of randomly selected neighbouring gas particles, $n_{\rm heat}$, by raising their temperatures by an amount $ \Delta T_{\rm heat}= 10^8$ K. In this way, the heated gas particles do not radiate away their thermal energy immediately but instead they drive supersonic outflows that are able to displace a large amount of gas far from the AGN themselves. 
A value of $\epsilon=0.015$ yields a good match to the $z=0$ relations between BH mass and stellar mass and velocity dispersion and the $z=0$ cosmic BH density.
On the scales of groups and clusters this is the most realistic simulation because it reproduces many observational data sets, such as the relations between X-ray luminosity, temperature, gas mass fraction, and SZ flux, as was shown by \citet{McCarthy10, McCarthy11} and \citet{LeBrun14}.
\item AGN 8.5: this model is identical to AGN 8.0 but with an increased AGN heating temperature of $\Delta T_{\rm heat}= 10^{8.5}$ K.  As per feedback event the same mass of gas is being heated in this model as in the fiducial AGN 8.0 model, more time is required for the BH to accrete sufficient energy to heat the gas by the higher temperature.  In practice, therefore, the duty cycles differ between the two models with the AGN 8.5 model having longer quiescent periods but a more energetic release of thermal energy in the surrounding medium for a given event.
 We note that $\Delta T_{\rm heat}$ can not be increased to arbitrarily high temperatures since this would lead to unrealistically long time periods between feedback episodes and would prevent self-regulation of the AGN feedback \citep[see][]{Booth09}. In a WMAP7 cosmology, \citet{LeBrun14} find that the AGN 8.0 and AGN 8.5 models effectively bracket the observed baryon fractions of local groups and clusters (see also Fig.~\ref{fig:mass_fgas}). We refer to \citet{LeBrun14} for the analysis of the BH population properties and for the BHs scaling relations showing that the BH formed in the simulations used in this paper are consistent with observational results and theoretical models. 
\item NOSN\_NOZCOOL:  in this simulation the SN feedback is removed and the gas cooling assumes primordial abundances.  No AGN feedback is included.
\item NOZCOOL: SN feedback is included but the gas cooling still assumes primordial abundances.   No AGN feedback is included.
\item WDENS: the SN feedback parameters depend on the local gas density of the star-forming particles from which the star particles that produce the SNe are formed. The initial wind velocity scales with density as $v_{\rm w} = 600 \kms  (n_{\rm H}/10^{-1} {\rm cm}^{-3})^{\frac{1}{6}}$ , which implies that $v_{\rm w}$  scales with the sound speed of the equation of state that we impose on the unresolved multiphase interstellar medium \citep{Schaye08}, and the mass loading factor as $\eta = 2  (n_{\rm H}/10^{-1} {\rm cm}^{-3})$. In this way the total amount of feedback energy per unit stellar mass is kept fixed. The higher wind velocity in dense gas results in a more efficient feedback in massive galaxies \citep{Haas12a}.   No AGN feedback is included.
\end{itemize}

A complete list of  simulations used in this paper, with detailed information on the  box size and resolution, is reported in Table \ref{tbl:sims}.

\subsection{Finding and matching haloes between simulations}
\label{Sec:match}

\begin{figure} \begin{center}
\includegraphics[width=1.0\columnwidth, trim=7mm 8mm 0mm 8mm]{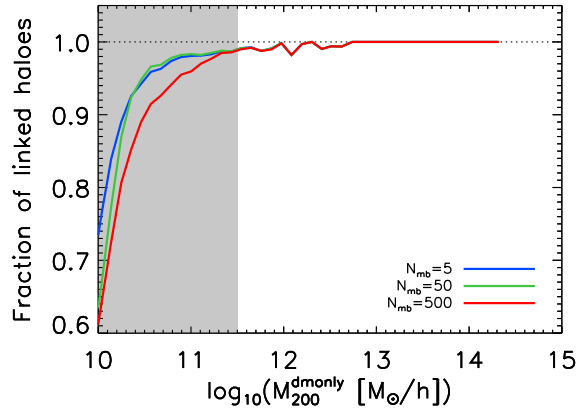}
\caption{The fraction of haloes that are successfully linked as a function of dark matter halo mass for FoF groups in DMONLY and REF. Each line shows what fraction of the FoF groups in DMONLY are linked to a FoF group in REF. Different colours are used for different values of the number of most-bound particles used to match haloes, the fiducial value being $N_\mathrm{mb}=50$. The grey shaded region is below the resolution limit. For haloes above this limit the fraction linked is very close to unity.}
\label{halofrac}
\end{center} \end{figure}

Haloes are identified in our simulations using the Friends-of-Friends algorithm combined with a spherical over-density algorithm centred on the minimum of the gravitational potential as implemented in \textsc{subfind} \citep{Springel01_subfind,Dolag09}. As every simulation from the OWLS project has identical initial conditions for a fixed box size, it is in principle possible to identify the same haloes in each simulation as these should contain mostly the same DM particles, which can be identified using their unique particle IDs. By linking haloes between simulations we can investigate how changes in physics influence the properties of a fixed sample of haloes. Specifically, we are able to examine how the halo mass changes from model to model.

The haloes linking procedure works as follows: for every halo in simulation A we flag the $N_\mathrm{mb}$ most-bound particles, meaning the particles with the highest absolute binding energy. Next, we locate these particles in the other simulations. If we find a halo in simulation B that contains at least 50\% of these flagged particles, a first link is made. The link is confirmed only if, by repeating the process starting from simulation B, the previous halo in simulation A is found.

Fig.~\ref{halofrac} shows the fraction of friends-of-friends (FoF) groups at $z=0$ that are successfully linked between DMONLY and REF as a function of dark matter halo mass. Different colours are used for different values of the number of most-bound particles used to match haloes, the fiducial value being $N_\mathrm{mb}=50$. For haloes above the resolution limit that we use for this work (see Appendix \ref{sec:res_test}), shown by the grey shaded region, the linked fraction is very close to unity. While this fraction insensitive to the value of $N_\mathrm{mb}$, using only a few particles to match haloes between simulations may lead to spurious matches, and, more importantly, increases the sensitivity to baryonic cooling and feedback. On the other hand, using values of $N_\mathrm{mb}$ that are too high means that even the most loosely bound particles of a halo are used up to relatively high masses, which leads to an even greater sensitivity to baryonic processes. The results for other pairs of simulations are similar, though the curves shift to the left when comparing simulations with baryons to each other, as haloes in these simulations are identified at lower masses. Matched haloes are not considered for our analysis if their mass $M_{200}$ in the DMONLY simulation is less than the mass resolution limit (see Appendix \ref{sec:res_test}).

\section{How baryons alter the masses of haloes}
\label{sec:result}

\subsection{Change in total mass between different realisations of the same halo}
\label{sec:diff}
\begin{figure*} \begin{center} \begin{tabular}{cc}
\includegraphics[width=1.0\columnwidth ]{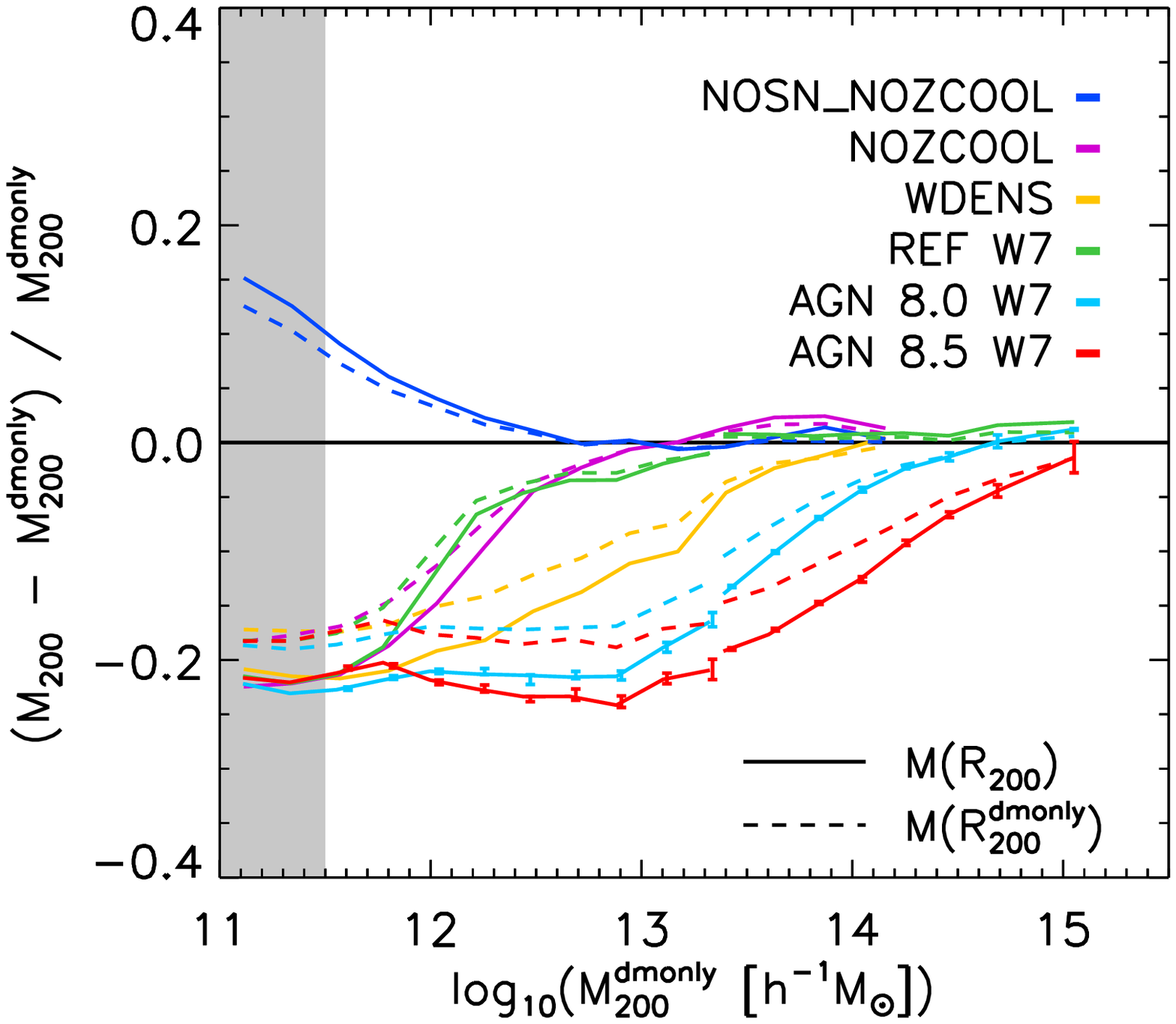} &   {\includegraphics[width=1.0\columnwidth  ]{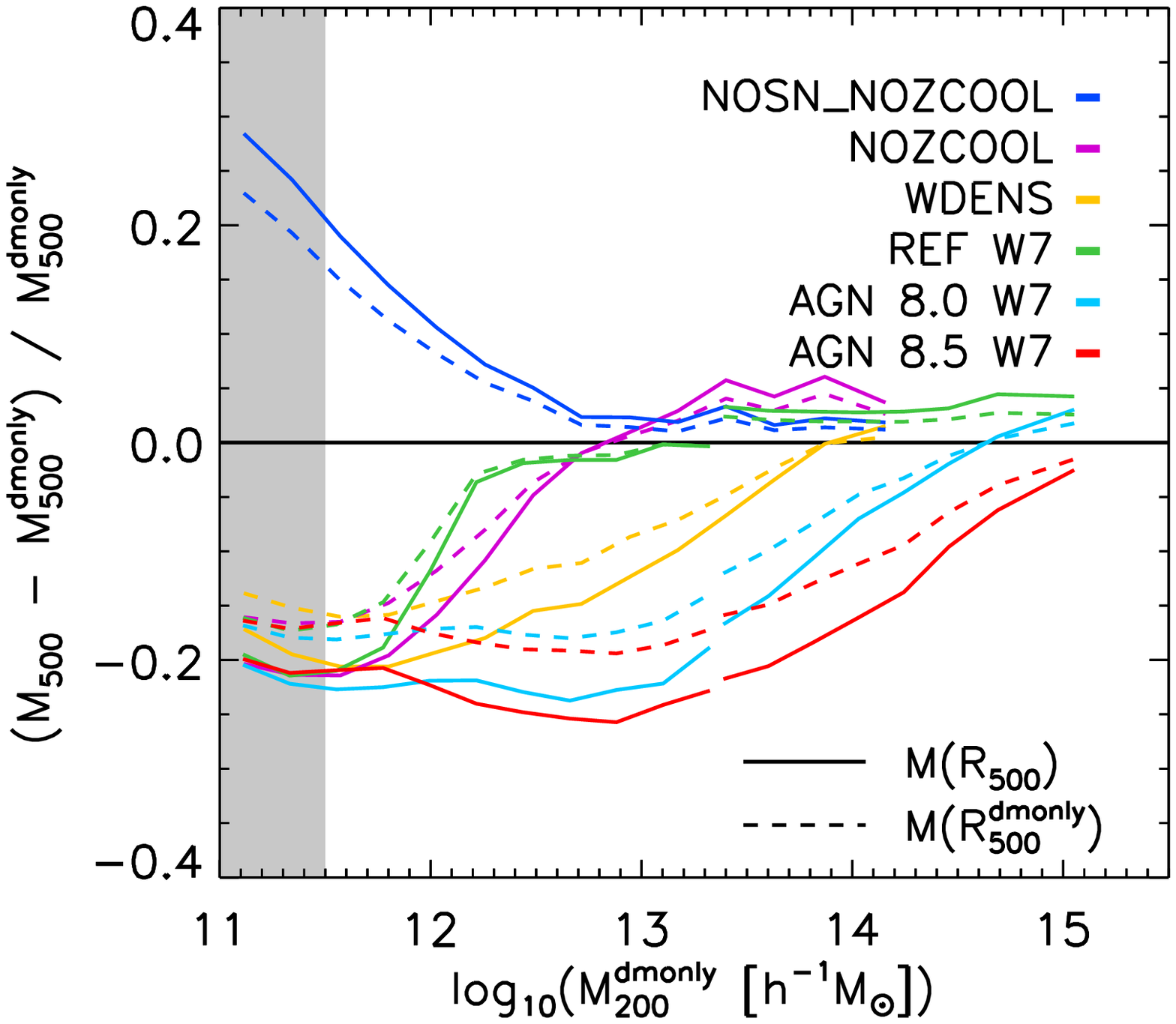}} \\
  {\includegraphics[width=1.0\columnwidth ]{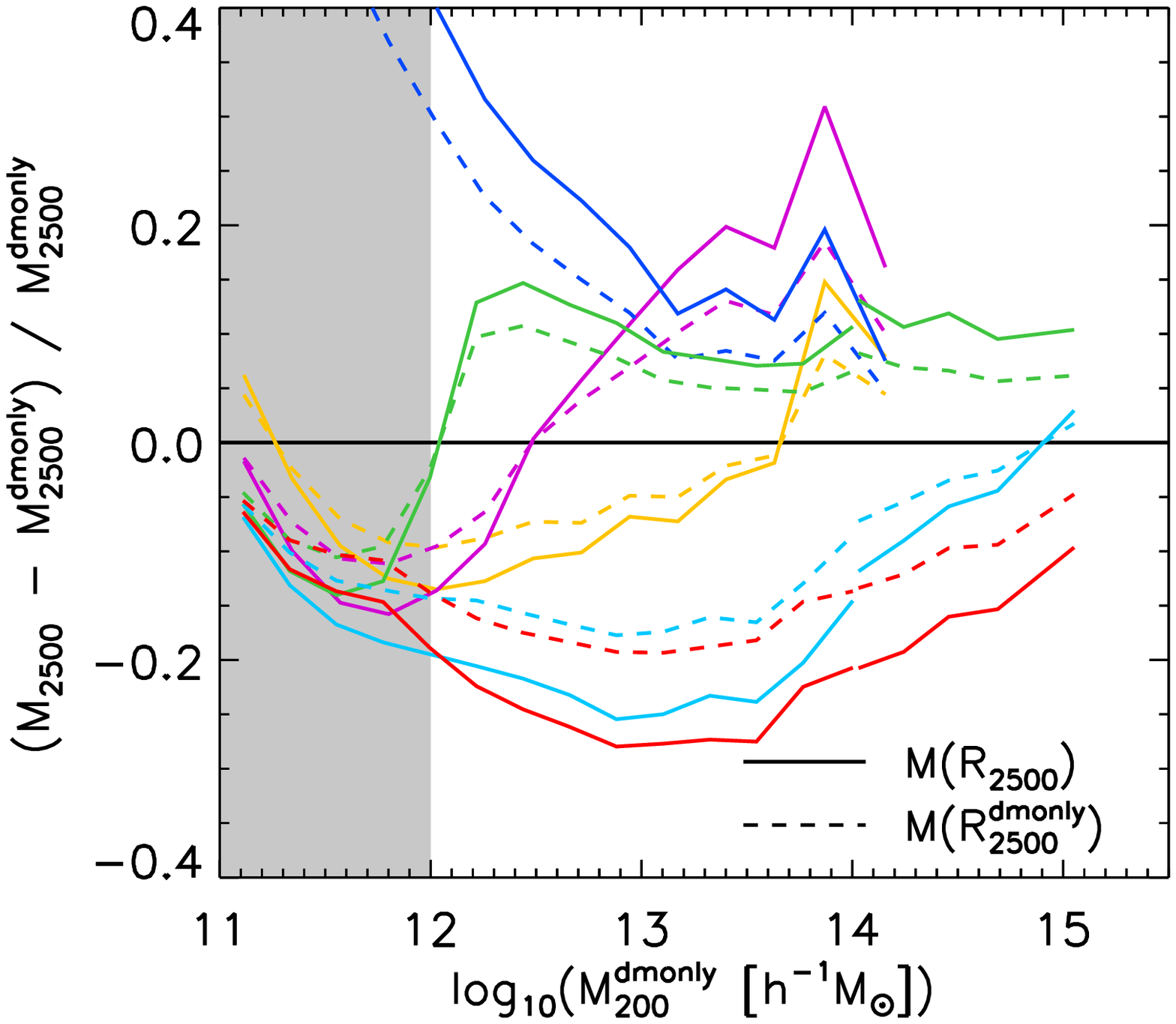}} & {\includegraphics[width=1.0\columnwidth  ]{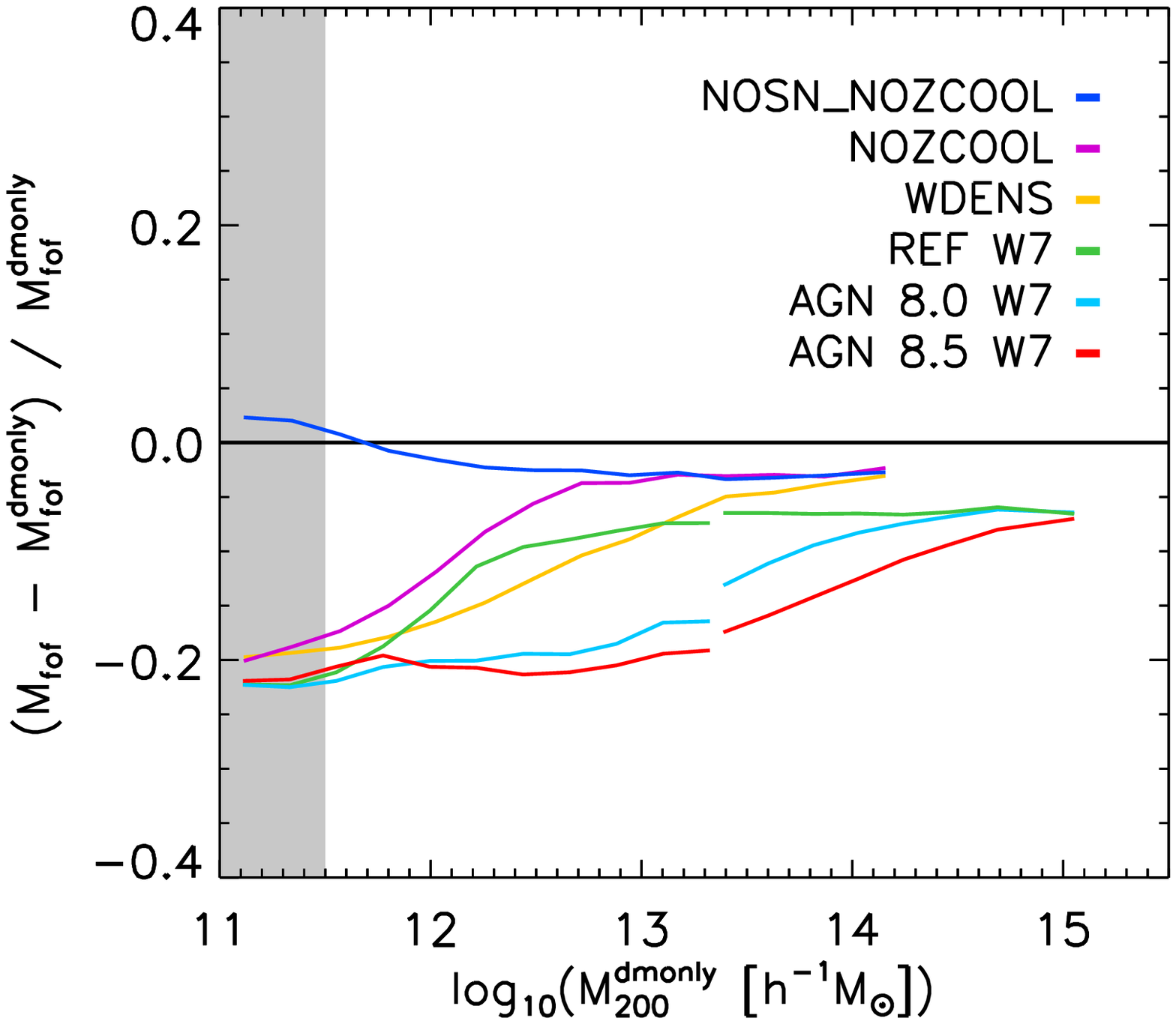}} \\
\end{tabular} \end{center}
\caption{The median relative differences in mass ($M_{200}$, $M_{500}$, $M_{2500}$ and friends-of-friends mass $M_{\rm{fof}}$) between matched haloes in the same mass bin at $z=0$ for all the different simulations. The dashed lines in the first three panels  show  the variation of the total mass inside a radius ($R_{\Delta}^\mathrm{dmonly}$) that is the same for any given different realisation of the same halo. Dark blue lines show the results for simulations when gas physics is introduced but not SN feedback or metal-line cooling (NOSN\_NOZCOOL). The effect of introducing SN feedback is shown by the purple lines (NOZCOOL). Green lines show the simulations with metal-line cooling and SN feedback (REF). The effect of increasing the efficiency of the SN feedback for massive haloes is shown by the yellow line (WDENS). Cyan lines represent the simulations that include SN feedback, metal-line cooling and AGN feedback with a heating temperature of $\Delta T_{\rm heat}= 10^{8.0}$ (AGN 8.0). The red lines show the version of the AGN simulation with higher heating temperature $\Delta T_{\rm heat}= 10^{8.5}$ (AGN 8.5), that produces a more efficient energy release on a longer duty cycle. All the simulations assuming the WMAP7 cosmology (REF, AGN 8.0 and AGN 8.5) are extended to higher masses by the use of cosmo-OWLS (done in a 400 $\hMpc$ box). The transition between the two box sizes happens when in a mass bin the number of haloes in the smaller box falls below 50 (10 for $M_{2500}$). The shaded grey region is below the resolution limit found in Appendix \ref{sec:res_test}. In addition, for the upper left panel, 2-sigma errors computed via bootstrapping are shown for the two AGN simulations.} 
\label{fig:diff200_z0}
\end{figure*} 

In this section we compare the masses of haloes that have been matched as described in Sec.~\ref{Sec:match}. Haloes are binned according to their mass in the DMONLY simulations ($M^{\mathrm{dmonly}}_{200}$). For each mass bin we plot the median value of the relative difference in mass with respect to the DMONLY realisation.  We limit our analysis to haloes with $M^{\mathrm{dmonly}}_{200} > 10^{11.5} \hMsun $ (which corresponds to a halo with about 600 particles; see Appendix \ref{sec:res_test}), use mass bins of width $ \delta \newl (M_{200}) = 0.25 $, and include all bins with at least 10 haloes. All the simulations assuming the WMAP7 cosmology (REF, AGN 8.0 and AGN 8.5) are extended to higher masses by the use of cosmo-OWLS (performed in a 400 $\hMpc$ box). The transition between the two box sizes happens when in a mass bin the number of haloes in the smaller box falls below a chosen value. Throughout the paper we adopt a value of 50, except for $M_{2500}$ for which we adopt 10. 

It is important to note that due to the spherical overdensity definition of masses $M_{\Delta}$ used in this work (and commonly adopted in the literature), a change in the total mass of haloes can be caused by a change in mass within the haloes as well as by change in their density profile.  
If the density profile changes, then so does the radius at which the value of the mean internal density reaches $\rho(r)=200\rho_{\mathrm{crit}}$. In this way a redistribution of matter inside the halo can change its total mass just by how $M_{200}^{\rm crit}$ is defined.   In order to isolate these two effects, we also show the variation of the total mass inside a given radius that is chosen to be the same in every realisation of the same halo among different simulations.

In Fig.~\ref{fig:diff200_z0} we show the relative difference in mass for haloes in different simulations with respect to the DMONLY simulation.  There are two different curves for every analysed simulation. The continuous curve represents the relative difference in the $M_{200}^{\mathrm{crit}}$ mass between the simulation with baryon physics and  the simulation without, namely $(M_{200} - M_{200}^{\mathrm{dmonly}}) / M_{200}^{\mathrm{dmonly}}$. In order to isolate the contraction or the ejection of baryons within a common radius, we also plot the quantity $[M(r<R_{200}^{\mathrm{dmonly}}) - M_{200}^{\mathrm{dmonly}}] / M_{200}^{\mathrm{dmonly}}$ (dashed lines). These curves give us insight into the different mass content of each halo realisation, because any differences are due only to the different amount of mass inside a common radius.

We focus first on the top left panel of Fig.~\ref{fig:diff200_z0}.

For the run without SN feedback, NOSN\_NOZCOOL (dark blue lines), it is clear from the low-mass end of Fig.~\ref{fig:diff200_z0} that the haloes are more dense than when they are simulated using only gravitationally interacting particles (DMONLY).  This is due to the absence of a mechanism that is able to heat the gas and prevent it from overcooling.  However, the mass difference becomes smaller with increasing halo mass, due to the increasing importance of gravitational shock heating which limits the cooling rate of the gas (e.g., \citealt{White78}). 

Next, we consider the addition of SN feedback with the simulation NOZCOOL (purple lines). Note that the cooling rate of the gas is still computed assuming primordial abundances.  For halo masses $ M_{200} \la  10^{12} \hMsun$, SN feedback leads to a  $\approx-20\%$ change in the mass with respect to the DMONLY simulation.  However, at higher masses the effects of SN feedback are minor and the difference with respect to the DMONLY simulation tends towards zero.  In other words, we find that SNe are incapable of ejecting gas from galaxy groups and clusters, consistent with previous studies (e.g., \citealt{Kravtsov05,Ettori06}).

If metal-line cooling is switched on (while retaining the SN feedback) as in the REF run (green lines), the predictions start to converge to the DMONLY simulation at slightly smaller halo masses.  This is because the feedback is less effective due to the increased cooling rate of the gas.  Note that for this simulation we use a WMAP7 cosmology in order to extend the dynamic range probed by our analysis by adding the cosmo-OWLS version of the REF simulation done in a 400 $\hMpc$ box.  We will explore the sensitivity of the results to changes in cosmology in Sec.~\ref{sec:cosmo} (they are minimal).

In the WDENS simulation (yellow lines), the parameters that regulate the wind scale with the local gas density, such that the wind velocity increases with density while the mass-loading decreases.  This has the net effect of increasing the effectiveness of the feedback in denser environments and high-mass haloes relative to the fixed-wind REF model.  Indeed, the WDENS curve has a similar shape to the REF curve, but the mass range over which the winds are effective is much more extended.

When AGN feedback is turned on (red and cyan lines) the picture changes. The decrease in halo mass extends to much higher masses than in models with SN feedback alone.  While the two AGN feedback models show the same qualitative behaviour, the model that invokes higher heating temperature (AGN 8.5) is able to extend the relative change in mass to higher masses due to the more effective energy release.  However, even AGN feedback is insufficient to significantly alter the total masses of the most massive galaxy clusters (with $M \sim 10^{15} \hMsun$).

Interestingly, in the highest mass bin in the AGN 8.0 simulation we obtain a slight {\it increase} in the mass of the halo simulated with AGN 8.0 with respect to the DMONLY case (the REF model also displays this behaviour).  This could be explained by the fact that the inclusion of baryons leads to more spherical haloes compared to the DMONLY simulation (e.g., \citealp{Bryan12}), and so the haloes in baryonic simulations tend to have more mass within a common radius due to this geometric effect.
\begin{figure} \begin{center} \begin{tabular}{c}
\includegraphics[width=0.9\columnwidth ]{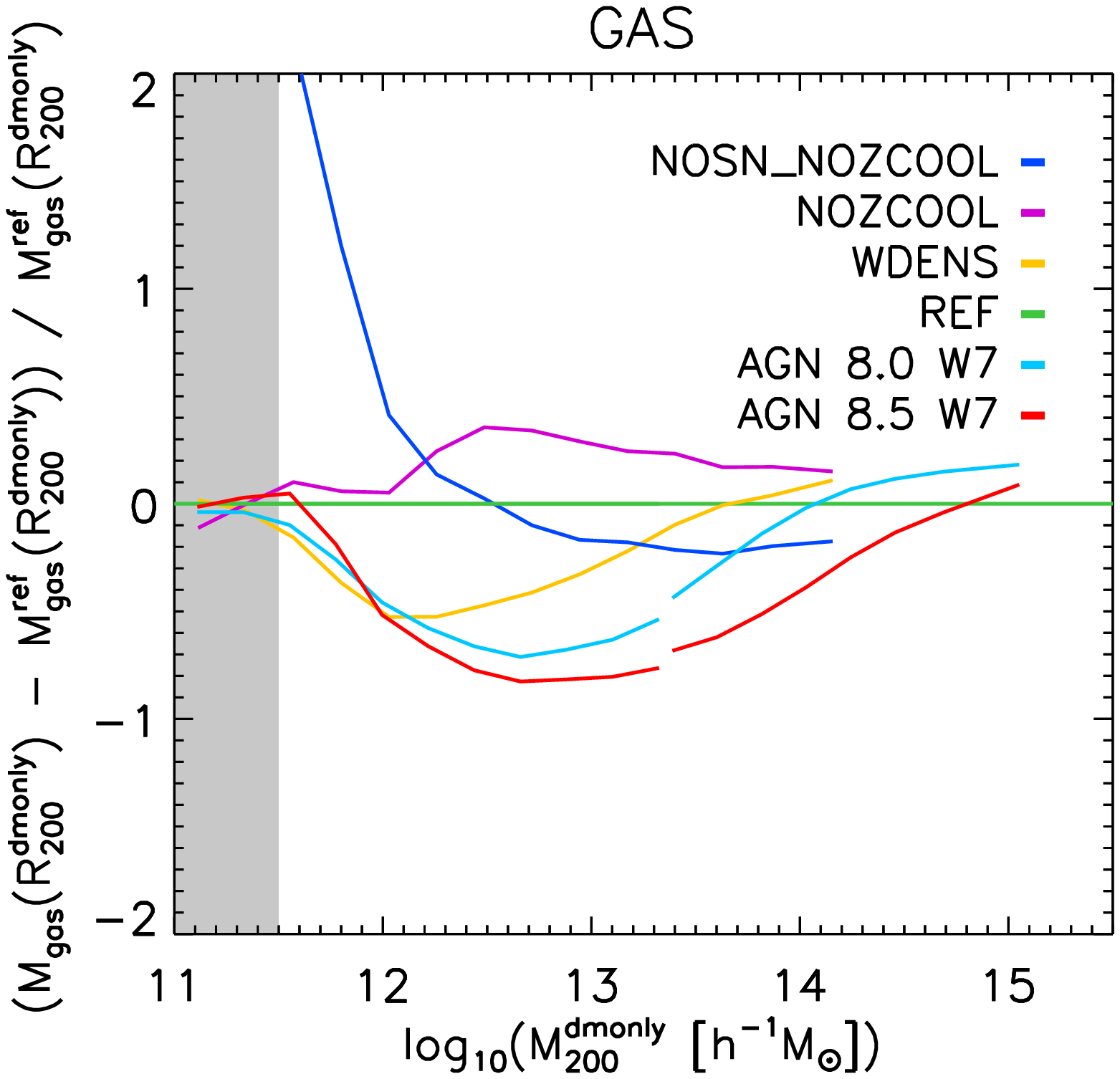} \\
   {\includegraphics[width=0.9\columnwidth  ]{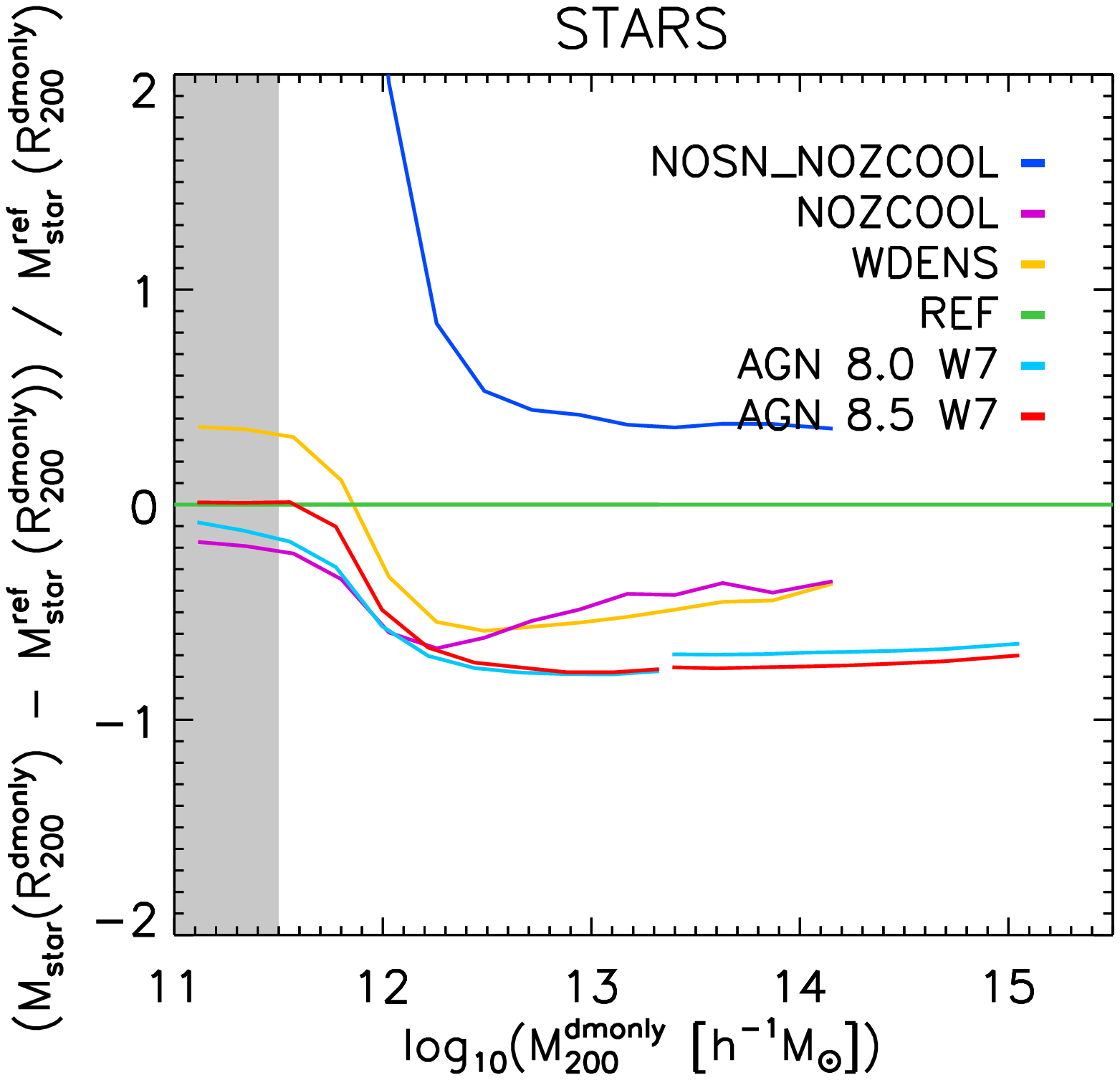}} \\
      {\includegraphics[width=0.9\columnwidth  ]{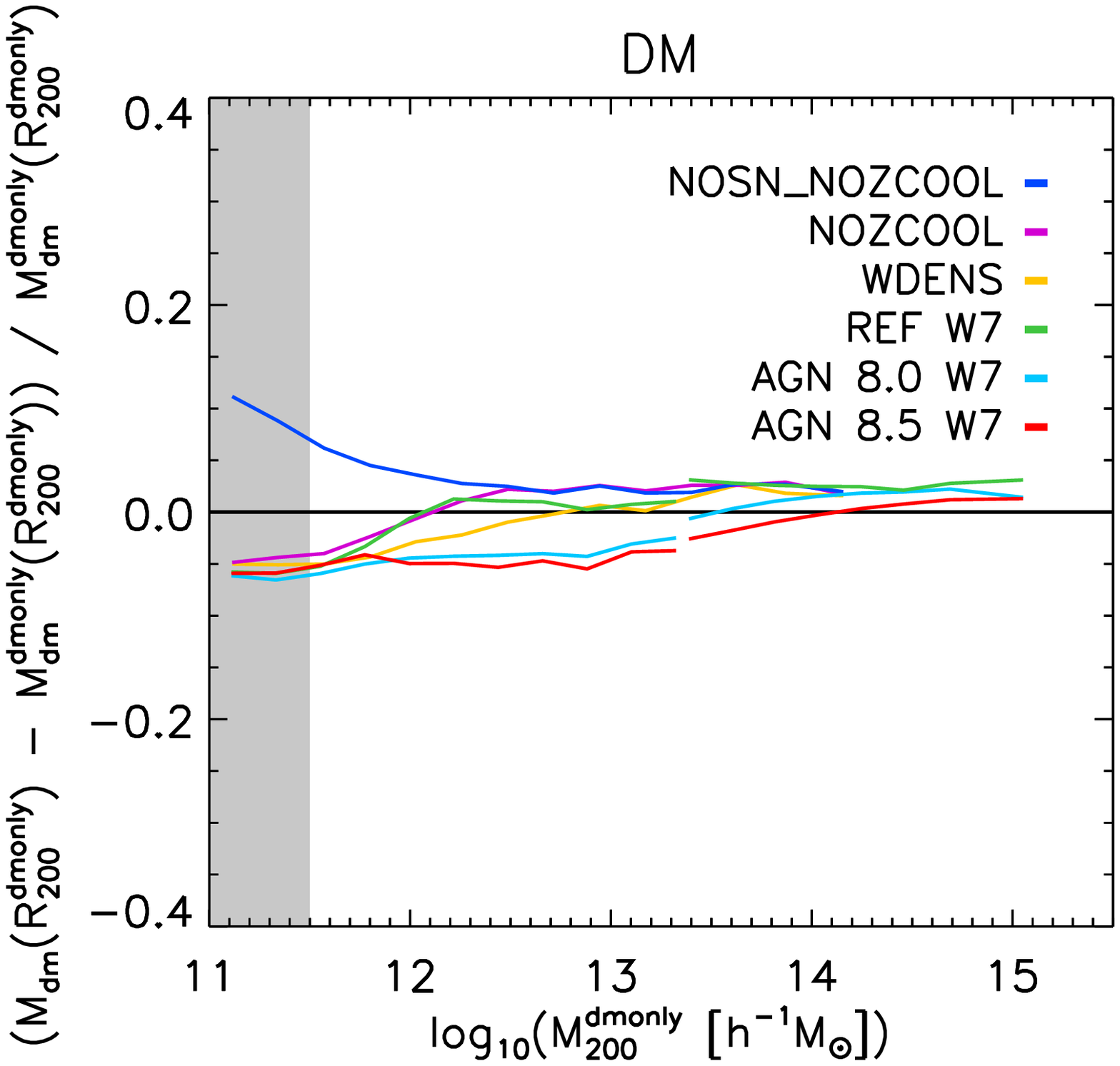}} \\
\end{tabular} \end{center}
\caption{Top panel: the relative difference in the median gas mass enclosed in $R^\mathrm{dmonly}_{200}$ with respect to the REF model in each simulation binned in $M^\mathrm{dmonly}_{200}$ mass. Middle panel: the relative difference in stellar mass in $R^\mathrm{dmonly}_{200}$ with respect to the REF model.
Bottom panel: the relative difference in dark matter mass in $R^\mathrm{dmonly}_{200}$ with respect to the DMONLY model, notice the change in scale. }
\label{fig:diff200_z0_comp}
\end{figure} 

So far we have discussed the top left panel in Fig.~\ref{fig:diff200_z0}. The top right ($M_{500}$) and the bottom left ($M_{2500}$) panels show the same analysis using higher over densities that probe the inner part of the haloes.
We note that for the simulations that include cooling but no feedback from AGN (NOSN\_NOZCOOL, NOZCOOL, REF, WDENS), the trend resembles the one seen in the $M_{200}$ panel, but with all the curves shifted upwards. This is because in the central region the overcooling dominates over the SN feedback (if any). In the case of central feedback (AGN) the difference with respect to the DMONLY simulation increases compared to the top left panel, since the feedback is more efficient in removing gas from the centre of the halo.

For completeness, in the bottom right panel in Fig.~\ref{fig:diff200_z0} we show the relative difference in the FoF mass ($M_{\mathrm{fof}}$). This mass represents the mass of all the particles that are linked together using the FoF algorithm and therefore within a DM isodensity contour. The linking of the particles in the FoF algorithm is done considering only DM particles (with linking length 0.2). Every baryonic particle is associated with its nearest DM particle. If this DM particle is in a FOF group, then the baryonic particle is assigned to the same FOF group.  Note that the relation will in general not converge to zero if the baryons do not have the same 3D spatial distribution as the DM.  However, we have tested that the DM component in the baryonic simulations converges to zero mass difference for high masses when the DMONLY simulation is rescaled to take into account the universal baryon fraction.

In order to evaluate the significance of the trends shown in Fig.~\ref{fig:diff200_z0}, we computed the errors on the median using the bootstrapping technique. These errors are shown in Fig.~\ref{fig:diff200_z0} for the two AGN models, but the amplitude is similar for the other mass definition and models. The errors are computed by taking the standard deviation of the distributions of the medians drawn from 1000 bootstrap realizations of the data in every mass bin. These errors are very small, which indicates that median values quoted in our analysis are robust. For the most massive bin the errors suggest that the median values are consistent with no change with respect to the DMONLY model.

The above analysis shows that important subgrid physics, particularly AGN feedback, can substantially alter the {\it total} masses of haloes, by up to $\approx20\%$ within $R_{200}$ (and by larger amounts within smaller radii).  This is suggestively close to (though slightly larger than) the universal baryon fraction, but as we will show later this does not imply that all of the baryons have been removed from the haloes (for an analysis of the gas and baryon fractions in these haloes we refer to Sec. \ref{sec:bar_frac}).  

In what follows immediately below, we quantify the effects of these subgrid processes on the stellar, gas, and dark matter masses separately.

\subsection{Change in baryon mass and back-reaction on dark matter}
\label{sec:diff_comp}

In the top and middle panels of Fig.~\ref{fig:diff200_z0_comp} we show the relative difference in the gas mass and stellar mass, respectively, within $R^\mathrm{dmonly}_{200}$ with respect to the REF model. The dark blue lines show the results for simulations that include gas physics but without SN feedback and metal-line cooling (NOSN\_NOZCOOL). For this model it is clearly visible that the mass in stars is much higher than the REF model (middle panel).  Again, this is due to the absence of a mechanism that prevents the gas from overcooling.  In terms of gas (top panel), at the low-mass end uninhibited cooling boosts the accretion of gas within $R^\mathrm{dmonly}_{200}$ compared to REF.  By contrast, at the high-mass end, where the SNe are not able to affect the total mass enclosed in REF, the amount of gas is less in the NOSN\_NOZCOOL model, since a larger fraction of the gas was converted into stars in REF (due to the increased cooling rate from metal lines).

When SN feedback is included but metal-line cooling remains off (NOZCOOL), a lower mass of stars is formed due to the increased effectiveness of SN feedback in the absence of metal-line cooling.  This leads to an increase in the mass of gas within $R^\mathrm{dmonly}_{200}$ with respect to the REF simulation. 

Inclusion of AGN feedback (AGN 8.0 and AGN 8.5) reduces the mass of stars and gas with respect to the REF model, the latter being due to ejection from the haloes.  The same qualitative effect is obtained at the high-mass end by varying the efficiency of SN feedback with the local gas density (WDENS).  On the other hand, at the low-mass end the SN feedback is less efficient at suppressing the star formation, increasing the amount of mass in stars with respect to the REF model.

In the bottom panel of Fig.~\ref{fig:diff200_z0_comp} we show the relative difference in the dark matter mass inside $R^\mathrm{dmonly}_{200}$ with respect to the DMONLY simulation due to baryon physics (notice the change in scale on the plot). At the low-mass end the effect of the overcooling and the consequent adiabatic contraction of the DM haloes is clearly visible for the simulation without SN feedback.  When SN feedback is introduced, the removal of gas expands the DM mass distribution, reducing the amount of DM mass that is present in the haloes. At the high-mass end gravity again becomes dominant and the adiabatic contraction of the DM distribution can only be offset by introducing AGN feedback.  However, even AGN feedback cannot prevent some slight contraction of the dark matter component on the scale of $R_{200}$ for the most massive haloes (clusters).

\subsection{Baryon fractions}
\label{sec:bar_frac}
\begin{figure} \begin{center} \begin{tabular}{c}
\includegraphics[width=0.9\columnwidth ]{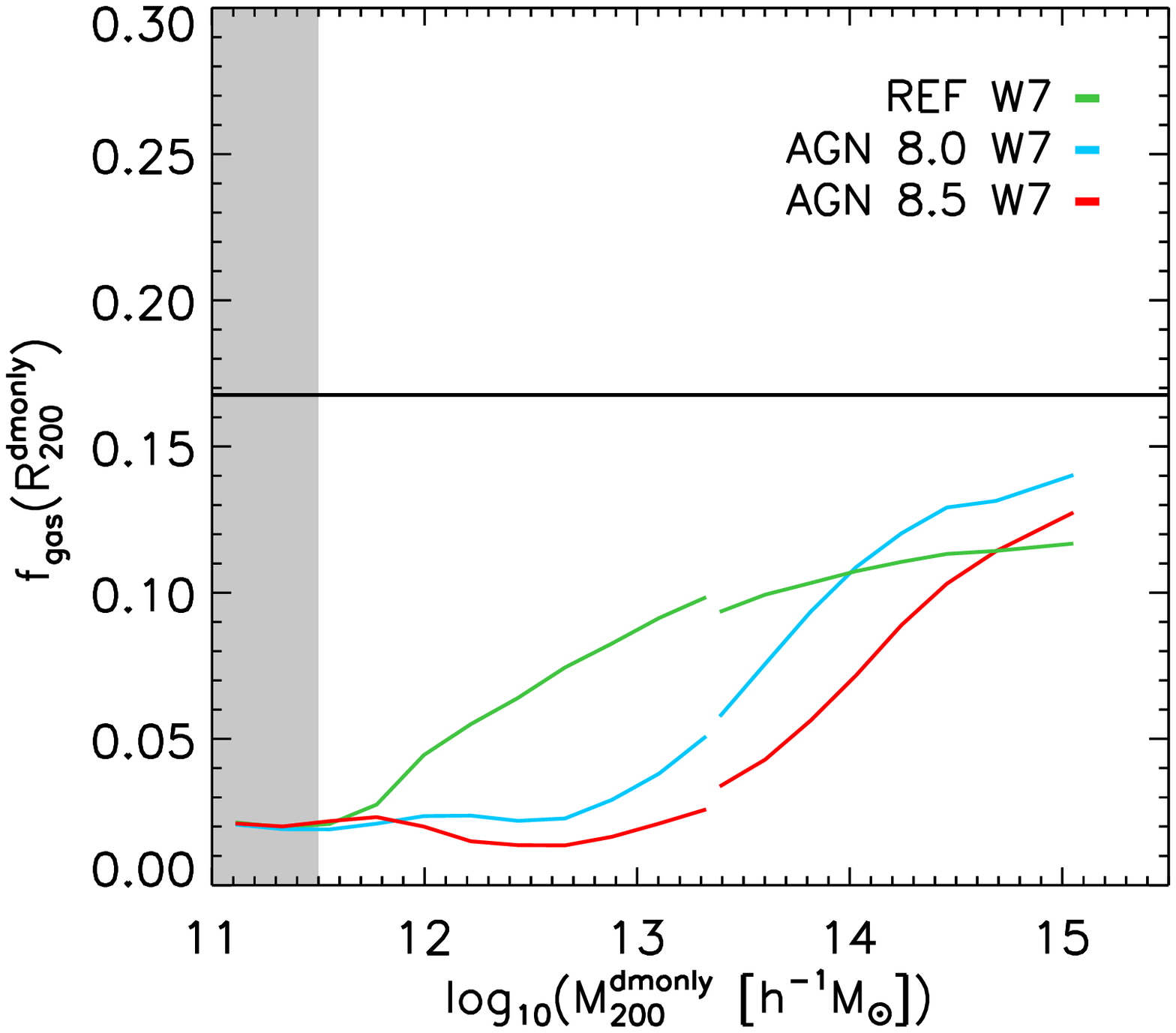} \\
   {\includegraphics[width=0.9\columnwidth  ]{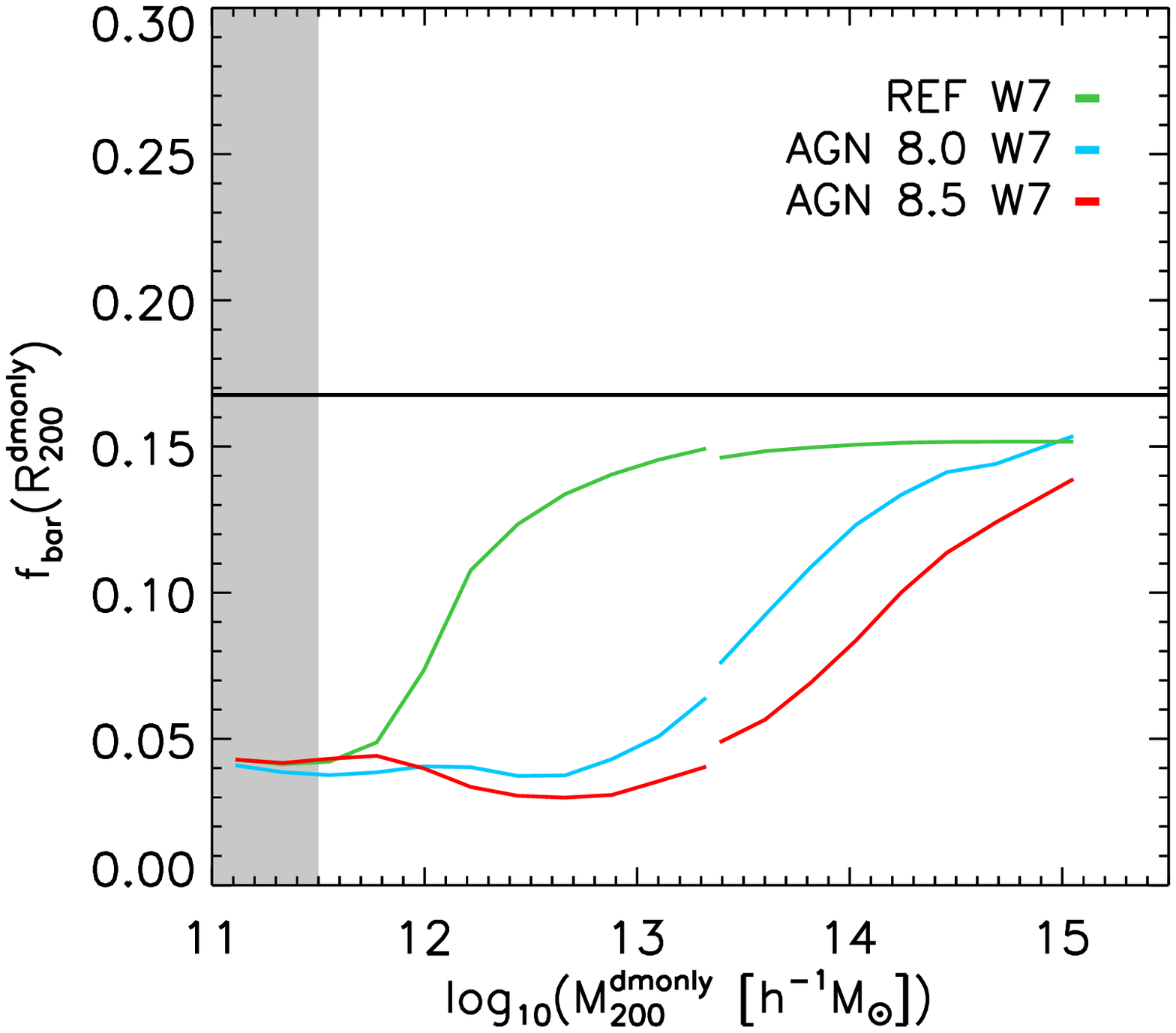}} \\
\end{tabular} \end{center}
\caption{The gas fraction (top panel) and baryon fraction (bottom panel) inside a sphere with a radius corresponding to the $R_{200}$ of the halo in the DMONLY simulation. The results are for the REF simulation (green lines) and the AGN simulations (cyan and red line). The continuous black line represent the universal baryon fraction for the WMAP7 cosmology.}
\label{fig:bar_frac}
\end{figure} 

In Sec.~\ref{sec:diff} we argued that the relative change in the total halo mass due to baryon physics is produced mainly by the ejection of baryons from the haloes, at least for models with efficient feedback.  The maximum magnitude of the effect is similar to the universal baryon fraction, which might naively suggest that most of the baryons have been ejected from the haloes.  Here we examine the baryon and gas mass fractions of haloes in the different simulations.

In Fig.~\ref{fig:bar_frac} we show the mass fraction in gas (top panel) and gas+stars (bottom panel) inside $R^\mathrm{dmonly}_{200}$ radius corresponding to $R_{200}$ of the same halo in the DMONLY simulation. We show the results for the REF simulation (green lines) and the AGN simulations (cyan and red lines). The continuous black line represents the universal baryon fraction for the WMAP7 cosmology.
It is clear that the amount of gas in the haloes is always less in the AGN simulations, with a minimum value that is $\approx15\%$ of the universal baryon fraction. Interestingly, for the highest mass bin in these simulations REF and the AGN have nearly the same gas mass but for very different reasons: in the REF simulation the gas is removed because it has been locked up in stars, while in the AGN simulations feedback from supermassive black holes expels much of the gas that would otherwise have been turned into stars.

Adding the stars to the gas we obtain the baryon fraction (bottom panel) and in this case the difference between the REF and the AGN models is striking.  The REF simulation produces haloes with baryon fractions close to the universal mean all the way down to low halo masses $M_{200}^{\mathrm{dmonly}}  \sim 10^{12.5} \hMsun$, in strong disagreement with observations of groups and clusters (e.g., \citealt{Budzynski14}).  The baryon fraction trend in the AGN simulations, by contrast, is very similar to that of the gas fraction, leading to a much better agreement with the observations, as shown by \cite{LeBrun14}.

From the above we see that even in the AGN models the haloes are not devoid of baryons.  This raises the question of how the total masses can be altered by up to $\approx20\%$ (for haloes with $M_{200}^{\mathrm{dmonly}}   \la 10^{13.5} \hMsun$).  The explanation is simply that the ejection of a large fraction of the gas expands the dark matter as well, as shown in the bottom panel of  Fig.~\ref{fig:diff200_z0_comp}.  It is the combination of gas ejection and dark matter expansion that causes a total mass change that is comparable to (slightly larger than) the universal baryon fraction.

\begin{figure} 
\begin{center} 
\includegraphics[width=0.9\columnwidth ]{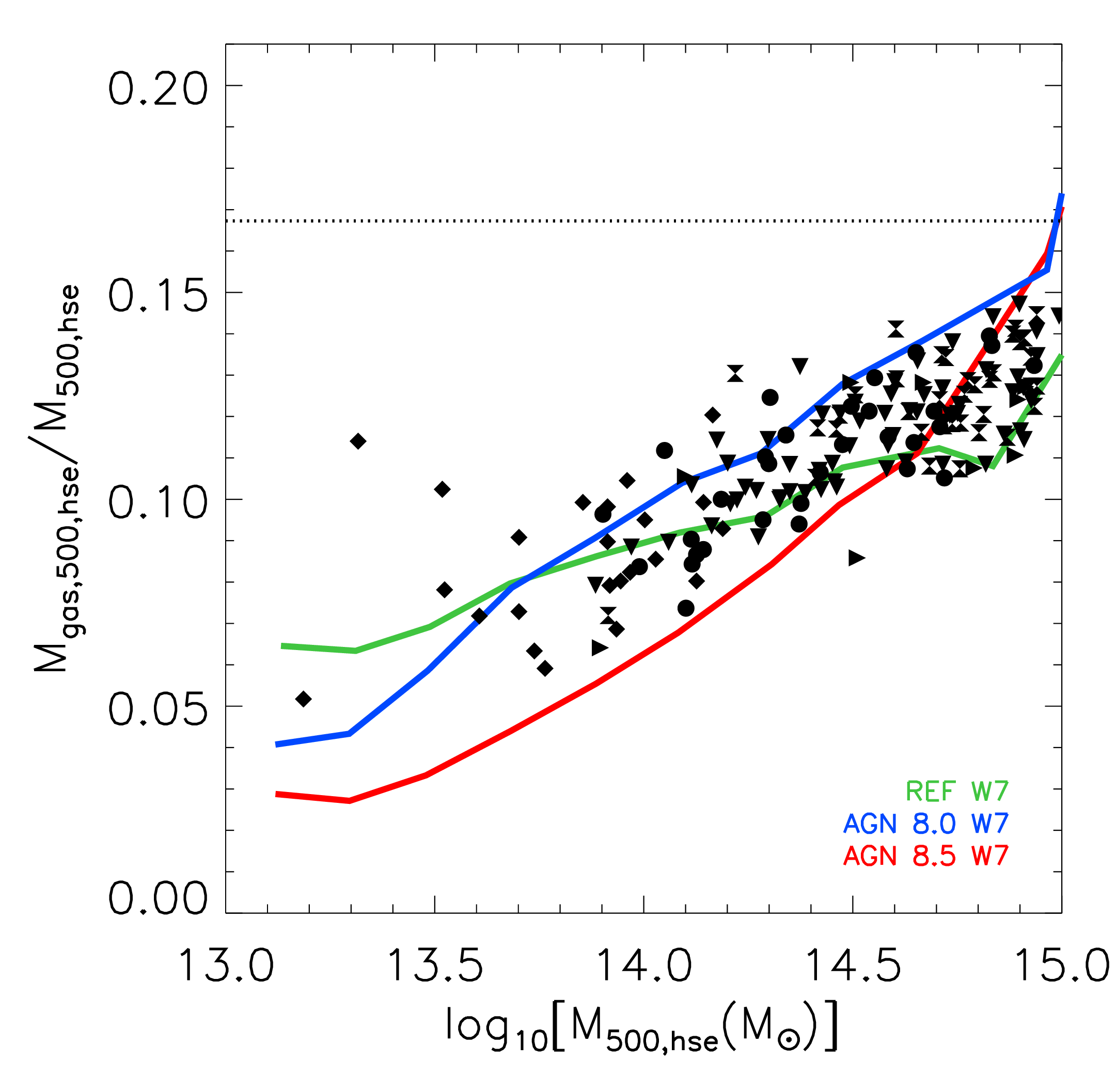} 
\end{center}
\caption{The gas mass fraction within $r_{500,\rm hse}$ as a function of $M_{500,\rm hse}$ at $z=0$. The filled black circles (clusters), right-facing triangles (clusters), downward triangles (clusters), hourglass (clusters) and diamonds (groups) represent the observational data of \citet{Pratt09}, \citet{Vikhlinin06}, \citet{Lin12}, \citet{Maughan08} and \citet{Sun09}, respectively. The coloured solid curves represent the median gas mass fraction--$M_{500,\rm hse}$ relations in bins of $M_{500,\rm hse}$ for the different simulations. The observed trend is approximately bracketed by the AGN 8.0 and the AGN 8.5 models for a WMAP7 cosmology.}

\label{fig:mass_fgas}
\end{figure}

It is important to compare our results for the simulated gas fraction with observations.  Hence, in Fig.~\ref{fig:mass_fgas}, we plot the gas mass fraction--$M_{500,\rm hse}$ relation at $z=0$ for the various simulations and compare to observations of individual X-ray-selected systems.  The gas mass fraction is measured within $r_{500,\rm hse}$.  For the simulations, we use the results from the synthetic X-ray analysis of \citet{LeBrun14} to `measure' the halo mass and gas mass fraction of the simulated systems, thus the 'hse' subscript in the masses indicates that they have been derived by a synthetic hydrostatic X-ray analysis.  As can be seen in Fig.~\ref{fig:mass_fgas}, the observed trend is approximately  bracketed by AGN 8.0 and AGN 8.5, at least up to masses of $10^{14.7} \hMsun$ in the WMAP7 cosmology (see \citealt{LeBrun14} for discussion). The REF model, which neglects AGN feedback, also yields reasonable gas mass fractions but they are achieved by overly efficient star formation as already shown in Fig.~\ref{fig:bar_frac}. Note that for $M_{500,\rm hse} < 10^{13.5} \Msun $the observational samples are likely biased.  We note also that the scatter about the median gas fraction trends (not shown here) is comparable to that in the observed relation (see \citealt{LeBrun14} for further discussion).

\subsection{Enclosed mass profiles}
\label{sec:mass_enc}

In this section we explore the variation of the total cumulative mass profile for haloes in different mass bins.  This is instructive for explaining the trends in the previous plots.  The reader who is interested only in the net effect on the halo mass function, may wish to skip to the next section.

\begin{figure*}
\begin{center}
\begin{tabular}{cc}
\includegraphics[width=0.5\textwidth ]{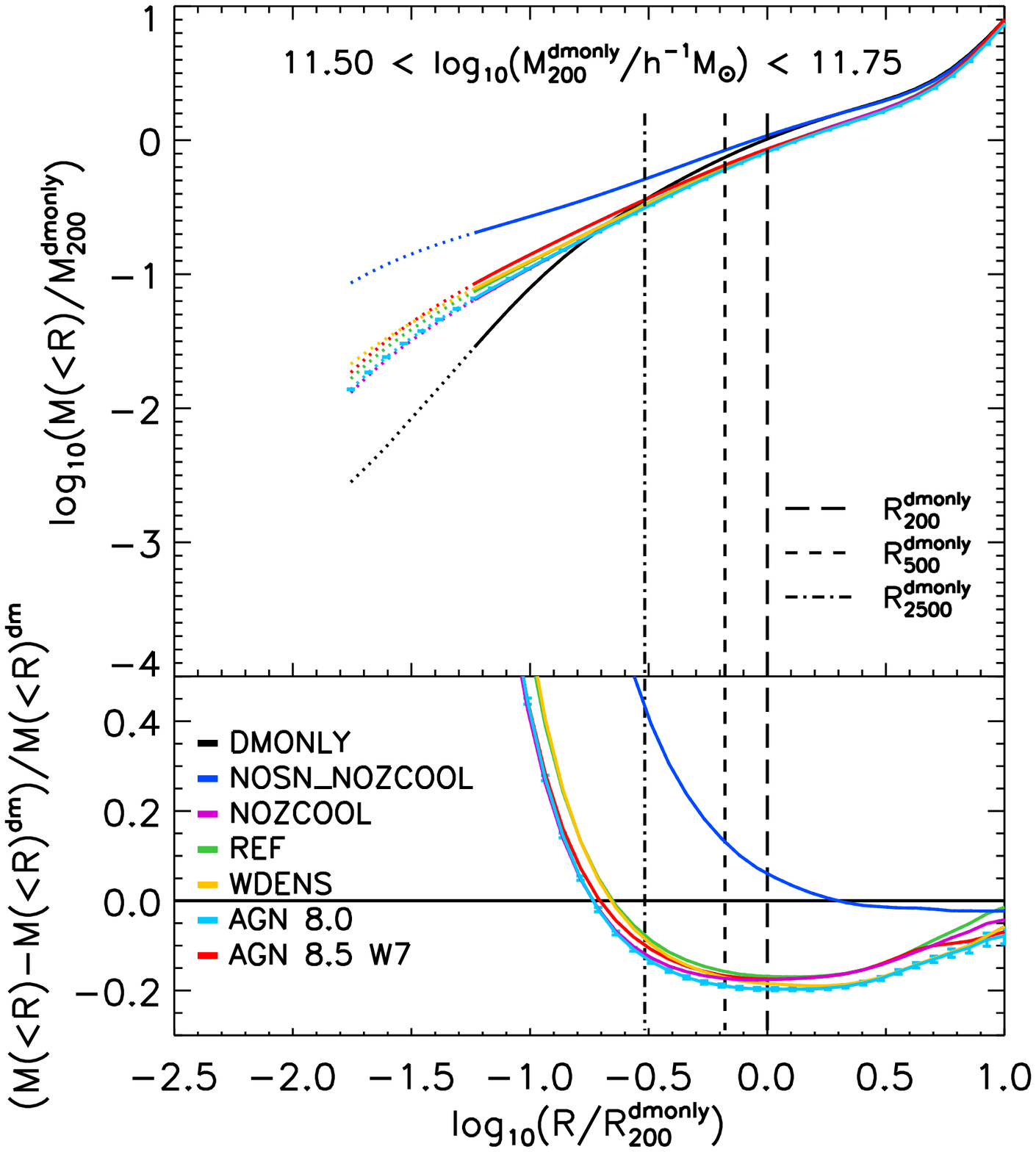} &  {\includegraphics[width=0.5\textwidth ]{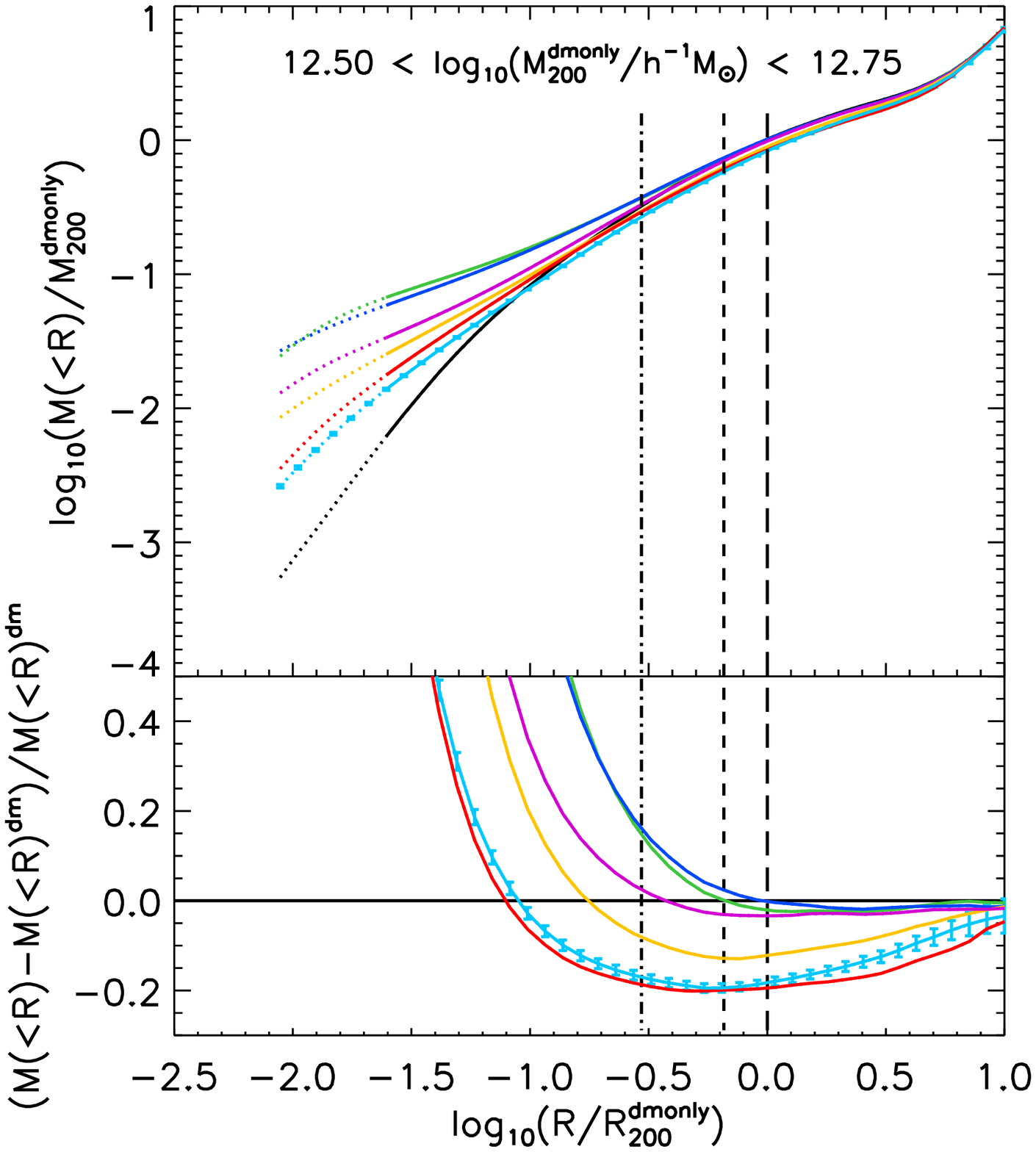}} \\
 {\includegraphics[width= 0.5\textwidth]{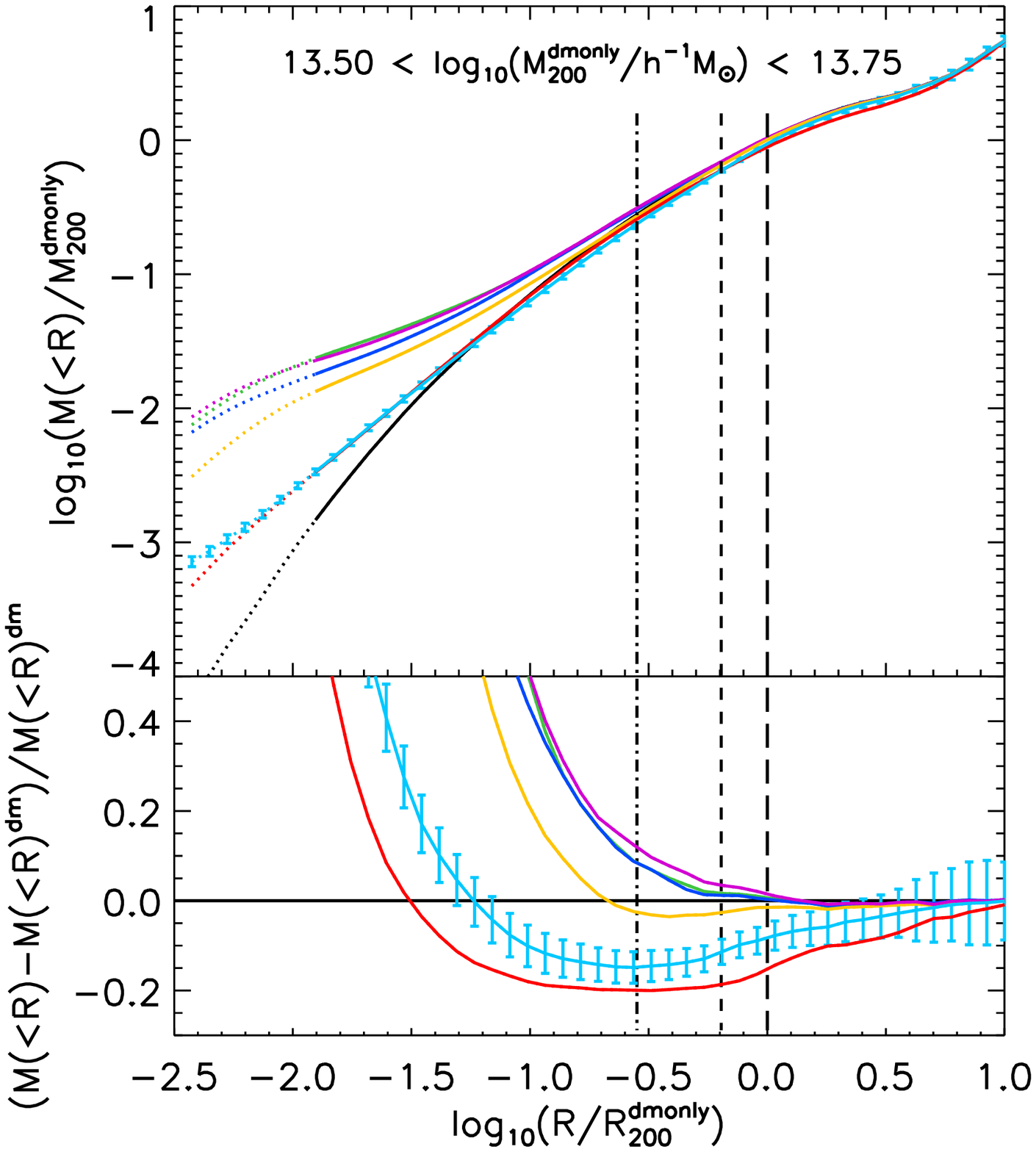}}& {\includegraphics[width=0.5\textwidth ]{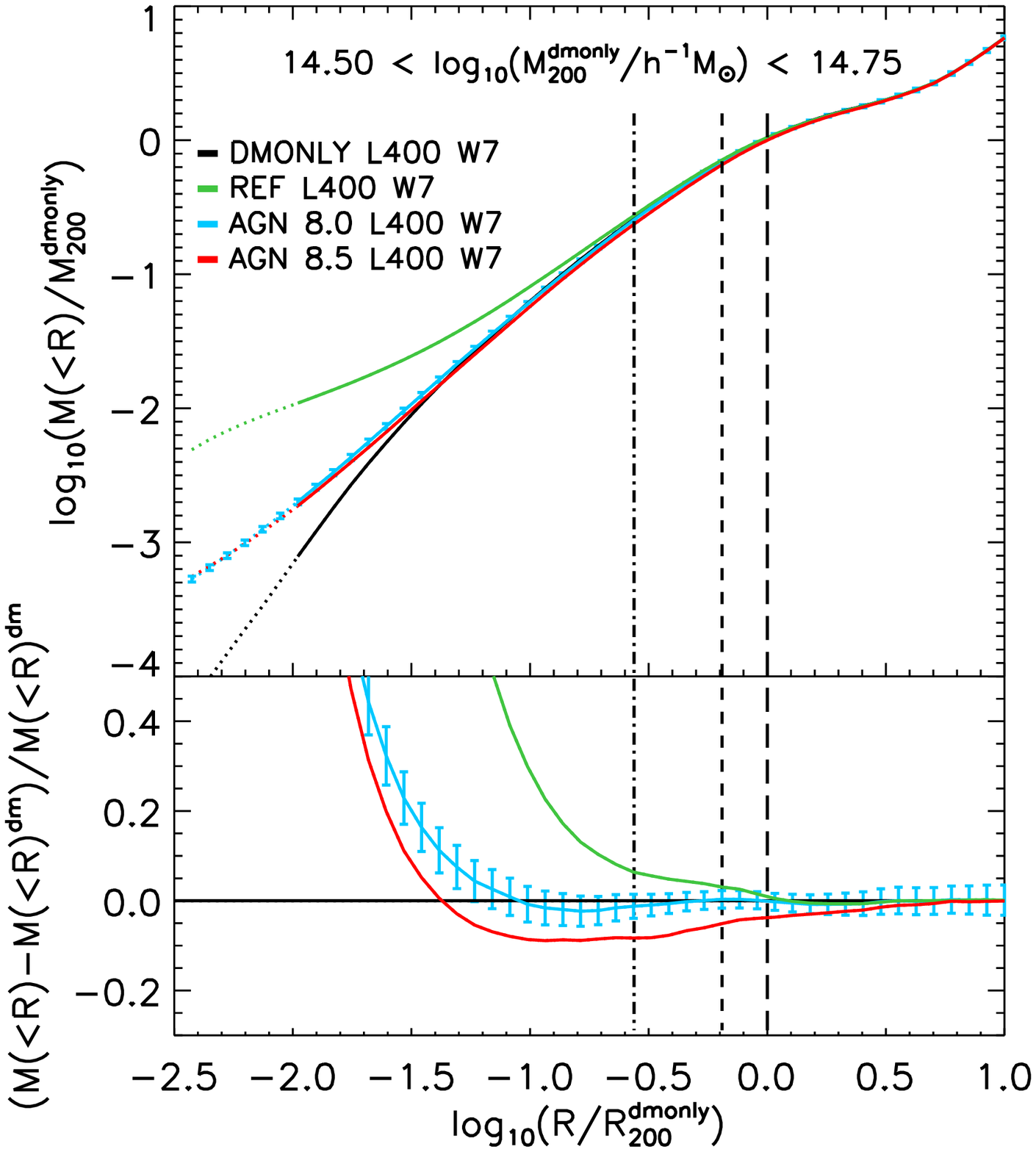}} \\
\end{tabular}
\end{center}
\caption{Median stacked profiles of the total enclosed mass as a function of radius in units of $R_{200}^{\rm dmonly}$. Each panel shows a different mass interval in  $M_{200}^{\mathrm{dmonly}}$. The long dashed vertical lines represent $R^{\mathrm{dmonly}}_{200}$ the radius of the halo in the DMONLY simulation, the dashed vertical lines represent the $R^{\mathrm{dmonly}}_{500}$ , and the dashed-dotted lines represent the $R^{\mathrm{dmonly}}_{2500}$. In the bottom part of every panel we show the relative difference between the curves of the simulations with baryons and the DMONLY simulations. 2-sigma errors from bootstrapping are shown for the AGN model. Every curve is dotted below three softening lengths and stopped at the softening length.}
\label{fig:massenc_profile}
\end{figure*}

For this analysis we use mass bins of size  $ \delta \newl M_{200}^{\mathrm{dmonly}} = 0.25 $, again selecting haloes by their mass in the DMONLY simulation.  For each simulation we take all the haloes in the mass bin and produce a median stacked total enclosed mass profile over the radial range $-2.5 \le \newl(R_{\rm min}/R_{200}^{\mathrm{dmonly}}) \le 1.0$. We use 47 bins over this radial range but plot only those bins which exceed the softening length of the simulation (below three softening lengths we use dotted lines).
The results are shown in Fig.~\ref{fig:massenc_profile}.  (Note that the variation of the mass enclosed at the radius equal to $R_{200}^{\mathrm{dmonly}}$ is what is shown by the dashed line in Fig.~\ref{fig:diff200_z0} for a given mass bin.)  The dot-dashed, short-dashed, and long-dashed vertical lines represent the median values of $R_{2500}^{\mathrm{dmonly}}$, $R_{500}^{\mathrm{dmonly}}$, and $R_{200}^{\mathrm{dmonly}}$, respectively.  Three different mass bins are shown for the $100\hMpc$ simulation box, and the last panel (bottom right) is taken from the $400\hMpc$ simulation. 

In the top left panel, we show the first mass bin, $10^{11.50}<M_{200}^{\mathrm{dmonly}} / [\hMsun] <10^{11.75}$. In this mass range it is clearly visible that in the inner regions (the central $\sim10\%$ of $R_{200}^{\mathrm{dmonly}}$) the baryonic component dominates in all the simulations.  However, at larger radii ($R \ga R_{2500}^{\mathrm{dmonly}}$) the mass enclosed becomes less than that in the case of DMONLY for simulations which include SN feedback (AGN, REF, WDENS, NOZCOOL).  Remarkably, at this mass scale convergence to the DMONLY result only occurs at very large radii of $R\ga5R_{200}^{\mathrm{dmonly}}$.

In the top right panel we consider the mass range $10^{12.50}<M_{200}^{\mathrm{dmonly}} / [\hMsun] <10^{12.75}$.  The trends are qualitatively similar to those in the top left panel, except that there is a much larger spread in the predictions of the models at small radii, due to the ineffectivness of SN feedback and the increasing importance of AGN feedback at these high masses.  Furthermore, models with no feedback or SN feedback alone converge to the DMONLY result at smaller radii ($R\sim1\mathrm{-}2R_{200}^{\mathrm{dmonly}}$) than models which also include AGN feedback ($R\ga5R_{200}^{\mathrm{dmonly}}$), as qualitatively expected.

The two bottom panels consider higher halo masses still:  $10^{13.50}<M_{200}^{\mathrm{dmonly}} / [\hMsun] <10^{13.75}$ (bottom left) and $10^{14.50}<M_{200}^{\mathrm{dmonly}} / [\hMsun] <10^{14.75}$ (bottom right).  Continuing the trends discussed above, the differences between the models are largest at small radii due to excessive overcooling in models without AGN feedback compared to those with it.  Note that even in massive galaxy clusters AGN feedback can noticeably alter the total mass distribution all the way out to $\sim R_{500}^{\mathrm{dmonly}}$.

We point out that the large-radii variation with respect to the DMONLY simulation (due to SN feedback at low masses and AGN feedback at high masses) does not necessarily mean that the baryons are ejected out to several times the virial radius, since much of the mass surrounding a given halo is in infalling galaxies which are also driving outflows and influencing their local environments.

\subsection{Evolution with redshift}
\label{sec:diff_evolution}

\begin{center} \begin{figure}
\includegraphics[width=\columnwidth ]{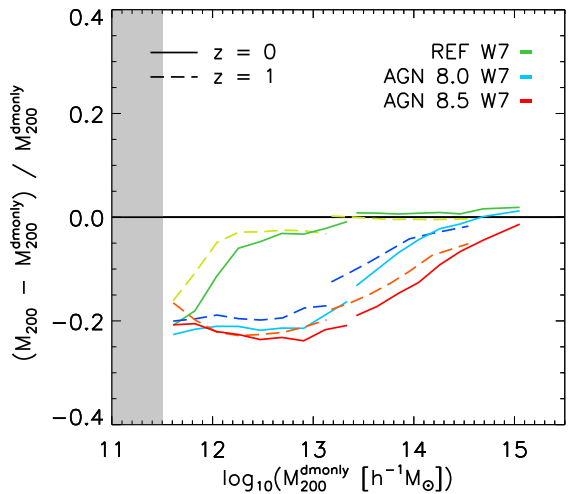}
\caption{Evolution with redshift of the relative difference in $M_{200}$ mass between simulations with baryon physics and  DMONLY. The results for redshift zero are represented by continuous lines, while the results for redshift one by long dashed lines.  }
\label{fig:diff_evo}
\end{figure} \end{center}

For completeness we explore the evolution of these trends with redshift, which is important since it is the evolution of the HMF that is the focus of upcoming cosmological studies.  In particular, in Fig.~\ref{fig:diff_evo} we compare the effects of SN and AGN feedback on the total mass of haloes at two different redshifts ($z=0,1$) using the three cosmo-OWLS simulations performed using a WMAP7 cosmology.  We show only these cases since they are representative of the evolution of the mass difference for all other simulations. 

In all the simulations it is clear that the absolute difference in the mass of haloes increases with time.  This is expected, since haloes are denser with increasing redshift and the the binding energy of a halo of fixed spherical overdensity mass therefore increases within increasing redshift.  Thus, more energy is required to alter the mass distribution of haloes (at fixed mass) at early times.  Overall, however, the difference in the trends between $z=0$ and $z=1$ for a given model is relatively minor.

We do not explore the difference in the relation for $z>1$ since our box is too small to provide a statistical sample of high-mass haloes at higher redshift.

\subsection{Effect of cosmology} \label{sec:cosmo}
\begin{center} \begin{figure}
\includegraphics[width=\columnwidth ]{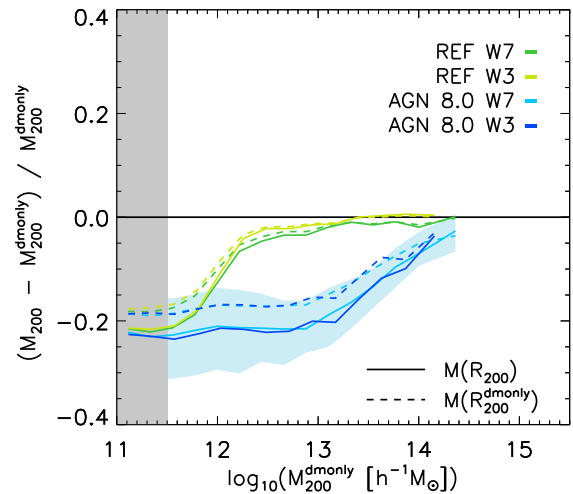}
\caption{Comparison of the effect of baryons on the mass of haloes for simulations with different cosmological parameters. The cyan curve shows the AGN 8.0 simulation in $100\hMpc$ box and using a WMAP3 cosmology. The blue line shows the same simulation done with the same box size but with a WMAP7 cosmology. For the REF simulations the green lines represents the WMAP7 variation and the light green the WMAP3 version. It is clear that the changes in the cosmological parameters between the WMAP3 and WMAP7 cosmologies do not lead to a significant change in the relative difference in halo mass. The light cyan band represent the area between the upper $84^{th}$ and the lower $16^{th}$ percentile for the AGN 8.0 W7 model.}
\label{fig:diff_cosmo}
\end{figure} \end{center}
We now test the sensitivity of our results to changes in the cosmological parameter values adopted in the simulations.

In Fig.~\ref{fig:diff_cosmo} we show the relative difference in halo mass due to baryonic effects using the AGN 8.0 model in a WMAP7 cosmology (cyan line) and compare it with the same model in WMAP3 cosmology (blue line).  The two simulations were run with the same resolution and with the same box size ($100\hMpc$). The two lines are fully consistent with each other.\footnote{The WMAP7 simulation line extends to slightly higher masses due to the larger value of $\sigma_8$ which yields a slightly larger number of high-mass haloes.}
The same test is presented for the REF simulations where the green line is the WMAP7 version and the light green represents the WMAP3 realization.  The convergence is excellent for both the total mass and the total mass enclosed in $R^\mathrm{dmonly}_{200}$ in this case.  (The slight difference between the two is plausibly due to the increased universal baryon fraction in the WMAP7 cosmology, which leads to slightly more gas cooling and slightly less efficient SN feedback.)
We also show the upper $84^{th}$ and the lower $16^{th}$ percentile around the median for the AGN 8.0 W7 model that is also representative for the scatter of the other simulations (for a more detailed description of the scatter see Sec.~\ref{Sec:fit_formula}).

We conclude that the analysis done on the variation of the mass of the halo due to baryon physics is largely independent of small variations in the input cosmological parameter values used.

\section{Analytic fitting formula for the change in halo mass} 
\label{Sec:fit_formula}

Combining the results obtained from the simulations in the $100 \hMpc$ box and the larger $400 \hMpc$ box, we are able to provide a fitting function that reproduces the median change in the mass of haloes due to baryon physics over four orders of magnitude in mass. We adopt the following functional form:
\begin{equation} 
\newl \left(\frac{M_{\Delta}}{M_{\Delta}^\mathrm{dmonly}}\right)  = A + \frac{B}{ 1 + \exp \left(- \frac{\newl (M_{\Delta}^\mathrm{dmonly})  + C}{D} \right)}, \label{eq:fit_func}
\end{equation}
This equation reaches the  constant value $A+B$ ($A$) in the high-mass limit and the constant value $A$ ($A+B$) in the low mass limit when the parameter $D>0$ ($D<0$).

We also provide a linear fitting function for the standard deviation of the change in halo mass at a given mass. The scatter is Gaussian when the logarithm of the difference is considered and is well approximated by the following fitting function:
\begin{equation} 
\sigma(\newl (M_{\Delta}^\mathrm{dmonly})) = E + F\newl (M_{\Delta}^\mathrm{dmonly})
\label{eq:fit_func_scatter}
\end{equation}  
The scatter is always a decreasing function of mass.  While the scatter is physical in origin, it is not strongly correlated with basic properties of the haloes like concentration.  We leave an examination of the origin of the scatter for future work.

When fitting Eqs.~\ref{eq:fit_func} and \ref{eq:fit_func_scatter}, we assign an error to each mass bin that is equal to the standard deviation of the distribution obtained from 1000 bootstrapped re-samplings.  

For the REF model we make use of three different simulations: a high resolution 50 $\hMpc$ version done using the WMAP3 cosmology that enables us to push the resolution limit down to $M_{200}^{\mathrm{dmonly}} / =10^{10.5} \hMsun $, the standard 100 $\hMpc$ WMAP7 version and the 400 $\hMpc$ box, also WMAP7, for the high-mass bins where the number of haloes in the smaller box falls below 50.  For the two simulations with AGN feedback the high-resolution version is not available, so we only combine the 100 and the 400 $\hMpc$ box both run with a WMAP7 cosmology.

In Table \ref{tbl:fit} we report the parameters of the fitting function (Eq.~\ref{eq:fit_func}) and the scatter (Eq.~\ref{eq:fit_func_scatter}), for different masses ($M_{200}$, $M_{500}$) and for the REF and the two AGN simulations. For other masses and redshifts the complete list of fitting parameters as well as the mean relations are available at \texttt{http://www.strw.leidenuniv.nl/MV14/}.

 We stress that this function is not meant to be extrapolated to masses lower than the mass resolution presented in this work, especially for the two simulations with AGN feedback. In fact, it is clear from the resolution tests that the relative difference in mass continues to increase in amplitude when the lowest mass regime is explored using simulations with higher resolution (Appendix \ref{fig:restest}). At the high-mass end the fitting parameters suggest that there is a constant offset value in the relative change in mass. However, we expect that the difference in mass converges towards no difference when the halo mass becomes sufficiently large. With this in mind, we also provide fitting functions that are constrained to asymptote to zero at high masses (given by the \emph{zero asint} tag in Table \ref{tbl:fit}).

\begin{center}
\begin{figure}
\includegraphics[width=\columnwidth ]{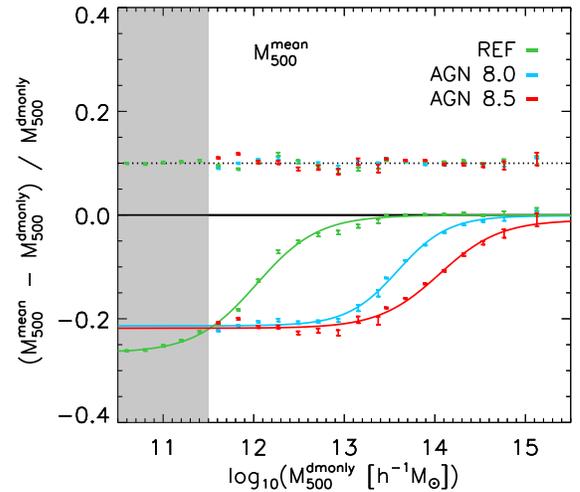}
\caption{Fitting function, constrained to asymptote to zero at high masses for the differences in  $M_{500}^{\rm mean}$ due to baryonic processes. The error bars show the $2-\sigma$ bootstrapped confidence interval. The lines show the best-fitting model using the function in Eq.~\ref{eq:fit_func}. The shaded region shows the resolution limit for the $100 \hMpc $ box, for the REF simulations the points below the resolution limit are taken from a high resolution version of the REF simulation. The error bars in the top half of the plot indicate the residuals between the fit and the points, shifted up by 0.1 for clarity.}  
\label{fig:fit}
\end{figure}
\end{center}

In Fig.~ \ref{fig:fit} we show the fitting function (constrained to asymptote to zero at high masses) for the differences in $M_{500}$ when the feedback processes are included. For the 100 $\hMpc$ simulation the unresolved regime is represented by a gray shaded region.  
It is clear that the fitting functions reproduce the trend of the simulations well.
Moreover, it is clearly visible that for the REF simulation the relative change in mass increases in amplitude towards the low-mass end when the simulation with higher resolution is used for the fitting.  This result is in agreement with the work of \citet{Sawala12} at the dwarf mass scale.

\begin{table} 
\begin{center}
\begin{tabular}{lclcccc}
\hline
Sim		             &        z       & Mass 				       & 		A (E)		& 		B (F)		& 		C		& 		D \\ \hline
\emph{AGN 8.0} &        0 &$ M_{200}$  &  -0.1080 &   0.1100 & -13.5861 &   0.2920 \\
\emph{zero asint} &        0 &$ M_{200}$  &  -0.1077 &   0.1077 & -13.5715 &   0.2786 \\
  &   &$ \sigma_{200}$  &   0.1294 &  -0.0082 &  &  \\
\emph{AGN 8.0} &        0 &$ M_{500}$  &  -0.1141 &   0.1232 & -13.6581 &   0.3114 \\
\emph{zero asint} &        0 &$ M_{500}$  &  -0.1133 &   0.1133 & -13.5947 &   0.2678 \\
\emph{ } &   &$ \sigma_{500}$  &   0.1166 &  -0.0069 &  &  \\
\emph{AGN 8.5} &        0 &$ M_{200}$  &  -0.0038 &  -0.1069 & -14.0424 &  -0.3398 \\
\emph{zero asint} &        0 &$ M_{200}$  &  -0.1109 &   0.1109 & -14.0745 &   0.3579 \\
\emph{ } &   &$ \sigma_{200}$  &   0.1104 &  -0.0064 &  &  \\
\emph{AGN 8.5} &        0 &$ M_{500}$  &  -0.0035 &  -0.1151 & -14.0871 &  -0.3333 \\
\emph{zero asint} &        0 &$ M_{500}$  &  -0.1187 &   0.1187 & -14.1132 &   0.3461 \\
\emph{ } &   &$ \sigma_{500}$  &   0.1073 &  -0.0060 &  &  \\
\emph{REF} &        0 &$ M_{200}$  &  -0.1385 &   0.1412 & -11.9307 &   0.3423 \\
\emph{zero asint} &        0 &$ M_{200}$  &  -0.1367 &   0.1367 & -11.9234 &   0.3148 \\
\emph{ } &   &$ \sigma_{200}$  &   0.1024 &  -0.0065 &  &  \\
\emph{REF} &        0 &$ M_{500}$  &  -0.1415 &   0.1517 & -11.7863 &   0.2791 \\
\emph{zero asint} &        0 &$ M_{500}$  &  -0.1366 &   0.1366 & -11.7623 &   0.2135 \\
\emph{ } &   &$ \sigma_{500}$  &   0.0910 &  -0.0053 &  &  \\
\hline
\end{tabular}
\caption{Fitting function parameters of Eq.~\ref{eq:fit_func} and for the scatter in Eq.~\ref{eq:fit_func_scatter}, for different simulations and different masses $M_{\Delta}^{\rm crit}$. The tag \emph{zero asint} gives the fitting function constrained to asymptote to zero at high masses} 
\label{tbl:fit} 
\end{center}
\end{table}

\section{Effects of baryons on the halo mass function}
\label{sec:hmf}
\begin{center} \begin{figure}
\includegraphics[width=\columnwidth ]{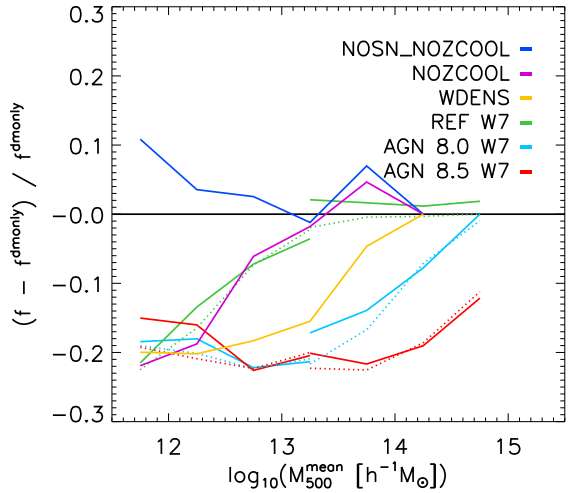} 
\caption{Relative difference in the HMF from simulations with baryons with respect to the DMONLY simulation. For the models for which the $400 \hMpc $  version is available we switch to the bigger box when the number of haloes per bin in the smaller box falls below 50. The dotted lines show the HMF computed by correcting the mass of every halo in the DMONLY simulation for the effects of baryon physics, using the fitting function provided in Sec.~\ref{Sec:fit_formula}. By applying this mass correction to the DMONLY haloes, we are able to reproduce the general trend of the change in HMF from simulations that explicitly include baryon physics. }
\label{fig:hmf_z0}
\end{figure}\end{center}
The halo mass function (HMF) gives the average number of haloes in a given mass range per unit volume. Usually the HMF is defined as:
\begin{equation} 
f \equiv  \frac{\mathrm{d} n}{ \mathrm{d}(\newl M)},  \label{eq:hmf_def}
\end{equation}
where $n$ is the number density of haloes and $M$ is the mass of haloes.

To calculate this function, we use the total halo mass $M^{\rm mean}_{500} $ enclosed within the radius $R^{\rm mean}_{500} $, defined as the radius within which the mean internal density reaches a value of $500 \times \rho_{\mathrm{mean}}$, where $\rho_{\mathrm{mean}}$ is the mean density of the Universe at that time.  We switch to using spherical overdensity masses defined with respect to the mean density as opposed to the critical density, since the former is more commonly used in mass function work (e.g., \citealt{Tinker08}).  Note that $M^{\rm mean}_{500} \approx 1.5 M^{\rm crit}_{500}$.

In Fig.~\ref{fig:hmf_z0}, we show the relative difference between the HMF from the simulations with baryons and the HMF obtained from the simulation with only gravitationally interacting particles (DMONLY).  The general trends in the relative difference in the HMFs are very similar to those in the relative change in mass (see Fig.~\ref{fig:diff200_z0}). This suggests that the major role in altering the HMF is played by the change in the masses of haloes rather  than by a change in the abundance of haloes. We also tested that the baryon physics does not change the abundance of haloes and found that the number of haloes that are matched between the simulations varies by less than one percent in every mass bin among the different simulations. 

We now test our ability to reconstruct the HMF of the simulations with baryon physics by starting from the DMONLY simulations and applying the change in halo mass. We first apply the mass change fitting functions presented in Section \ref{Sec:fit_formula} to every halo in the DMONLY simulation and then recompute the HMFs assuming that the scatter in the mass change does not play an important role. We use the fitting functions that are constrained to asymptote to zero for high masses, since even a small constant change in halo mass at very high masses can produce a non-converging result in the HMF due to its steepness at the high-mass end.

The results are shown by the dotted lines in Fig.~\ref{fig:hmf_z0}.  
It is immediately apparent that the dotted lines reproduce the general trend of the change in the HMF correctly. The main difference with the HMF of the REF simulation at the high-mass end is due to adopting the fitting function that goes to zero for higher masses when it is clear that in the REF simulation there is a constant positive offset in the halo masses with respect to the DMONLY case.  Both the AGN cases are reproduced quite well by this change in mass, especially at high masses.

Thus, the change in the mass of haloes is responsible for the differences in the HMFs introduced by baryon physics, and we have shown by applying the fitting function for the change in halo masses, we are able to reproduce the HMFs of simulations with baryons starting from a simulation with only dark matter.
Below we will apply the change in mass due to baryon physics to a generic HMF fitting formula obtained from N-body simulations that can be applied to different cosmologies.

\subsection{Analytic fitting formula for the halo mass function} 
\label{Sec:hmf_fit_formula}

Assuming that the change in the total halo mass is insensitive to small changes in the cosmological parameters, as the analysis in Sec. \ref{sec:cosmo} suggests, we can apply the mass correction (Eq.~\ref{eq:fit_func}) to a theoretical prescription for the HMF.

We use the formalism of \cite{Tinker08} for the theoretical mass function.  In order to obtain the linear variance over a certain mass scale, $\sigma(M)$, we assume a linear power spectrum, we apply the transfer function as presented in \cite{Eisenstein98} and we assume a top-hat window function in real space. We use the fitting parameters, calibrated using DM-only simulations, for the normalized version of the fitting function $g(\sigma)$ as presented in the appendix of \cite{Tinker08} and used in \cite{Tinker10}.
Here we just summarize the equations for calculating the halo mass function:
\begin{equation}
\label{e.dndsigma}
\frac{dn}{dM} = g(\sigma)\,\frac{\bar{\rho}_{\rm m}}{M}\frac{d\ln \sigma^{-1}}{dM}.
\end{equation}
Here, the function $g(\sigma)$ is expected to be universal to the
changes in redshift and cosmology and is parametrised as
\begin{equation}
\label{e.fsig}
 g(\sigma) = B\left[\left(\frac{\sigma}{e}\right)^{-d} + \sigma^{-f}\right]e^{-g/\sigma^2}
\end{equation}
and normalized as follows:
\begin{equation}
\label{e.norm}
\int g(\sigma)\,d\,\ln \sigma^{-1} = 1.
\end{equation}
The expression for $\sigma$ is
\begin{equation}
\sigma^2 = \frac{1}{2\pi}\int P(k)\hat{W}^2(kR)k^2dk,
\end{equation}
where $P(k)$ is the linear matter power spectrum as a function of wavenumber $k$, and $\hat{W}$ is the Fourier transform of the real-space top-hat window function of radius $R$.

Since the scatter (in the mass change due to baryons) does not play a major role in shaping the HMF, we apply only the median change in mass relation presented in Sec.~\ref{Sec:fit_formula} in order to get the HMF with the effects of AGN feedback included, according to:
\begin{equation} 
\frac{dn}{dM} (M^{\rm agn}_{\Delta}) = \left( \frac{dn}{dM} \right)^{\rm dmonly} ({M^{\rm dmonly}_{\Delta}}{(M^{\rm agn}_{\Delta})}).
\label{eq:new_hmf}
\end{equation}

Moreover, we can fit the relative difference in the halo mass function using the functional form already used in the previous section, providing in this way an easy to use correction function. The fitting function becomes: 

\begin{equation} 
\newl \left(\frac{f_{\rm agn}}{f_{\rm dmonly}}\right)  = A + \frac{B}{ 1 + \exp \left(- \frac{\newl (M_{\Delta}^\mathrm{dmonly})  + C}{D} \right)}, 
\label{eq:fit_func_hmf}
\end{equation}
where $f$ is defined in Eq.~\ref{eq:hmf_def}. The parameters of the fitting function in Eq.~\ref{eq:fit_func_hmf} are presented in Table \ref{tbl:fit_hmf}.

\begin{center}
\begin{figure}
\includegraphics[width=\columnwidth ]{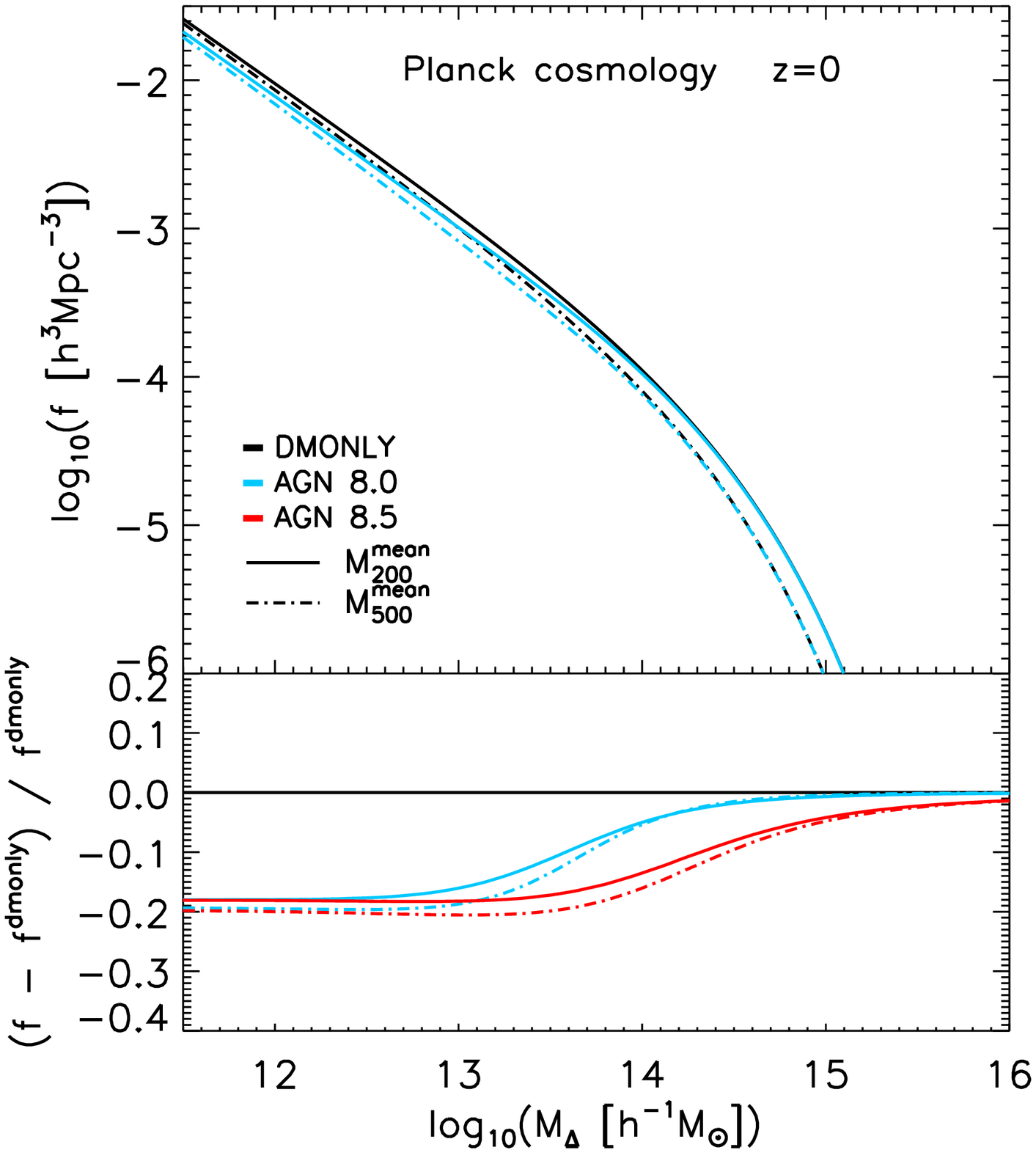}
\caption{In the top panel the black lines show the halo mass function computed using the fitting formula from \citet{Tinker08} fot Planck cosmological parameters \citep{Planck13}. The cyan curves show the Tinker HMF but corrected for the change in mass calibrated on the AGN 8.0 simulations. In the bottom panel we show the relative difference with respect to the uncorrected Tinker HMF, and we also add the relative change when AGN 8.5 is used (red). The different lines are: $M_{200}^{\rm mean}$ (continuous line) and $M_{500}^{\rm mean}$ (dash-dotted line).}
\label{fig:fit_hmf_200_500}
\end{figure}
\end{center}

As an example, we show in Fig.~\ref{fig:fit_hmf_200_500} the $z=0$ HMF for $M_{200}^{\rm mean}$ (continuous line) and $M_{500}^{\rm mean}$ (dash-dotted line).  The black lines correspond to the uncorrected DMONLY HMF using the fitting formula from \citet{Tinker08} with Planck best-fitting cosmological parameters (Planck+WP+highL+BAO; \citealt{Planck13}).   The cyan lines show the mass-corrected HMF according to the results from one AGN simulation (AGN 8.0), using the fitting functions that are asymptote to zero at high masses. In the bottom panel we show the relative difference between the curves also, adding in red the results from the AGN simulation with a higher heating temperature (AGN 8.5).

\begin{center}
\begin{figure}
\includegraphics[width=\columnwidth ]{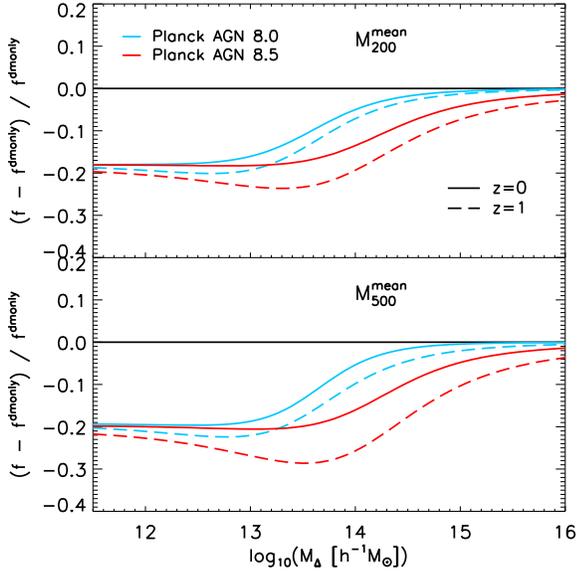}
\caption{Relative differences of the halo mass function when the correction for the change in mass is applied. In the top panel are shown the results for $ M_{200}^{\rm mean}$ at redshifts $z=0$ (continuous line) and at $z=1$ (dashed line). Results in the bottom panel refer to $ M_{500}^{\rm mean}$.
}
\label{fig:fit_hmf_200_500_z_evo}
\end{figure}
\end{center}

In Fig.~\ref{fig:fit_hmf_200_500_z_evo} we show the relative difference in the halo mass function for two different redshifts.  Interestingly, the relative difference in the HMF at $z=1$ is larger than at $z=0$, a trend that is opposite with respect to the trend in the relative mass change (see Fig.~\ref{fig:diff_evo}).  This is due to the rapid evolution of the HMF between these two redshifts.  The HMF at $z=1$ is steeper than it is at $z=0$ and, even though the relative change in halo mass is smaller, this results in a larger change in the HMF at $z=1$.


\begin{table} 
\begin{center}
\begin{tabular}{lclcccc}

\hline
Sim			& z & Mass 				& 		A 		& 		B		& 		C		& 		D \\ \hline
\emph{REF} &        0 &$ M_{200}^{\rm mean}$  &  -0.1155 &   0.1155 & -12.0603 &   0.4230 \\
\emph{REF} &        0 &$ M_{500}^{\rm mean}$  &  -0.1203 &   0.1203 & -11.9864 &   0.3487 \\
\emph{AGN 8.0} &        0 &$ M_{200}^{\rm mean}$  &  -0.0872 &   0.0872 & -13.6339 &   0.3509 \\
\emph{AGN 8.0} &        0 &$ M_{500}^{\rm mean}$  &  -0.0942 &   0.0942 & -13.7063 &   0.2717 \\
\emph{AGN 8.5} &        0 &$ M_{200}^{\rm mean}$  &  -0.0881 &   0.0881 & -14.4100 &   0.4280 \\
\emph{AGN 8.5} &        0 &$ M_{500}^{\rm mean}$  &  -0.0976 &   0.0976 & -14.4808 &   0.3795 \\ \hline
\emph{REF} &        1 &$ M_{200}^{\rm mean}$  &  -0.1232 &   0.1232 & -11.7513 &   0.3970 \\
\emph{REF} &        1 &$ M_{500}^{\rm mean}$  &  -0.1174 &   0.1174 & -11.5540 &   0.1876 \\
\emph{AGN 8.0} &        1 &$ M_{200}^{\rm mean}$  &  -0.0903 &   0.0903 & -13.8505 &   0.2978 \\
\emph{AGN 8.0} &        1 &$ M_{500}^{\rm mean}$  &  -0.0993 &   0.0993 & -14.0034 &   0.2979 \\
\emph{AGN 8.5} &        1 &$ M_{200}^{\rm mean}$  &  -0.0995 &   0.0995 & -14.7619 &   0.3603 \\
\emph{AGN 8.5} &        1 &$ M_{500}^{\rm mean}$  &  -0.1149 &   0.1149 & -14.9524 &   0.3177 \\
\hline
\end{tabular}
\caption{Fitting formula parameters of Eq.~\ref{eq:fit_func_hmf} calculated using the Planck cosmology, for different simulations and different mass definitions.} 
\label{tbl:fit_hmf} 
\end{center}
\end{table}

\subsection{Implications for cluster number counts}
\label{sec:dndz}

As discussed in Section 1, the number density of high-mass haloes and its evolution with redshift are sensitive to a number of fundamental cosmological parameters that control the growth rate of structure.  There are numerous ongoing and planned surveys whose main aim is to constrain these parameters by counting the number of high-mass systems on the sky.  As we have shown, however, the mass function is also sensitive to the (subgrid) physics of galaxy formation.  Here we propagate these effects to show the impact on the predicted number of massive haloes.

We define a cluster to have a mass of $M^{\rm mean}_{500} \ge 10^{14} \hMsun$ and compute the number of haloes above this mass limit at a given time for a comoving volume element.  

More specifically, we calculate the function:
\begin{equation} {\cal N}(z)=\frac{dV}{dz}\;\int_{M_{1}}^{M_{2}} n(M,z)dM,
\end{equation}
where $n(M,z)$ represents the HMF and $dV/dz$ is the comoving volume element, which in a flat universe takes the form:
\begin{equation} 
\frac{dV}{dz} =4\pi r^{2}(z)\frac{dr}{dz}(z),
\end{equation} 
with $r(z)$ denoting the comoving radial distance out to redshift $z$:
\begin{equation} 
r(z) = \frac{c}{H_0}\int_{0}^{z} \frac{dz'}{E(z')}. 
\end{equation} 

We account for the effect of baryon physics by using the HMF modified to include the change in the mass of haloes as described in the previous section.  We examine only the AGN models since we know that SN feedback alone is insufficient to change the masses of haloes in this mass range (and also leads to significant overcooling in disagreement with observations).  For every redshift we integrate the halo mass function at that redshift corrected by the effect on the total mass at redshift zero, in this way we assume that the relative change in mass does not vary with redshift.A more consistent way would be to interpolate the parameters of the available fitting functions for every given redshift where we calculate the number counts. Since the difference in the fitting functions at the three redshifts for which we compute them, z=0, 0.5, 1, are small, and because we have many more haloes at z=0 than at higher redshift, we assume that the change in mass does not vary with redshift.
Moreover, we neglect issues having to do with survey completeness and selection effects that can alter the cluster number counts.  These issues clearly need to be properly addressed when comparing to a specific survey.

\begin{center}
\begin{figure}
\includegraphics[width=\columnwidth ]{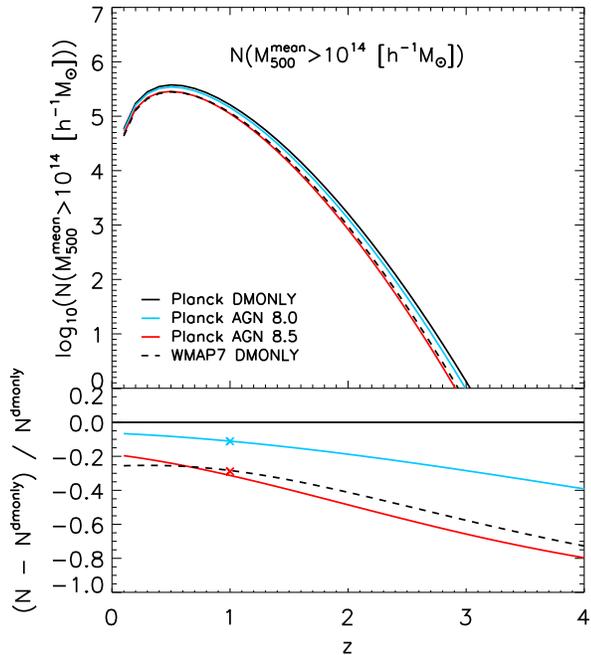}
\caption{The comoving number density of haloes more massive than $ M^{\rm mean}_{500} =10^{14}\hMsun$ as a function of redshift for the WMAP7 (dashed black line) and the Planck cosmology (continuous black line) predicted by \citet{Tinker08}, i.e. for a DM only universe. We also show the effect of using the HMF corrected for the change in mass of the haloes calibrated on the AGN 8.0 simulation (continuous cyan line) and for AGN 8.5 simulation (continuous red line). For this analysis we assume that the relative change in mass does not vary with redshift. In the bottom panel we show the relative difference of the functions with respect to the Planck DMONLY case. The two crosses at $z=1$ represent the values obtained by applying the change in mass fitting function at the same redshift. It is clear from this figure that the effect on the masses of the haloes due to baryon physics can produce a difference of the same order as the one produced by interesting variations of the cosmological parameters.
}
\label{fig:dndz}
\end{figure}
\end{center}

In Fig.~\ref{fig:dndz} we show the comoving number density of haloes more massive than $ M^{\rm mean}_{500} = 10^{14} \hMsun$ as a function of redshift for the WMAP7 (dashed black line) and the Planck cosmology (continuous black line) predicted by \cite{Tinker08}, i.e. for a DM only universe. We also show the effect of using the HMF corrected for the change in mass of the haloes calibrated on the AGN 8.0 (continuous cyan line) and AGN 8.5 (continuous red line) simulations.  Here we see that the inclusion of baryon physics (AGN feedback in particular) can lead to an effect that is of the same order as a change between the best-fit WMAP7 and Planck cosmologies. Thus, for precision cosmological work it is clear that the effects of baryon physics on the HMF must be modelled.

Finally, we tested what the effect is of assuming that the change in mass does not vary with redshift by adding in Fig.~\ref{fig:dndz} two points, one for each model, that indicate the cluster number count at $z=1$ using the mass correction predicted by the simulations for the same redshift. The difference between the points and the lines represents the error introduced by assuming a change in the mass relation that does not vary with redshift. This difference is indeed very small, thus validating our initial assumption. It is important to note that the fitting function at $z=0$ was based on 982 haloes with masses $ M^{\rm mean}_{500} >10^{14}\hMsun$ for the AGN 8.0 simulation (862 for the AGN 8.5), while for the fitting function at z=1 only 43 haloes are considered for the same mass range. This means that the fitting function at z=0 is better constrained than the fitting function at $z=1$. 

It is important to note, however, that the magnitude of the effect is quite sensitive to the mass limit used to define a cluster.  Here we have adopted a mass limit of $ M^{\rm mean}_{500} =10^{14}\hMsun$, which is roughly comparable to that of surveys such as XMM-XXL, XCS, and GAMA.  Surveys such as REFLEX II and Planck, which have mass limits that are a factor of several higher than this, will be considerably less sensitive to the effects of baryons on the HMF. When a higher mass limit of $ M^{\rm mean}_{500} =10^{15}\hMsun$ is used, correcting the masses of the cluster according to the results for AGN 8.5 reduces the cluster counts only by 10\% at $z=1$ (1\% for AGN 8.0). Instead, for the same mass limit and redshift, the change in the cosmological parameters from Planck to WMAP7 has a much bigger impact, reducing the cluster counts by 50\%. Note however that we only have 6 haloes with masses $ M^{\rm mean}_{500} > 10^{15}\hMsun$ at z=0 for AGN 8.5 (7 for AGN 8.0) to constrain the  behaviour, in the high-mass regime, of the fitting function that we apply to derive the corrected cluster number counts. Because of the poor statistics for very high mass haloes in our simulation box the results for this higher mass limit are less robust than the correction to the cluster number counts with a limit mass of $ M^{\rm mean}_{500} =10^{14}\hMsun$. Nonetheless, this result suggests that the impact of baryons on the cluster number counts becomes less severe when a higher mass limit is adopted.

\section{Comparison with previous studies}
\label{sec:discussion}

There have been several recent works examining the inclusion of baryons on the HMF.  In this section we compare our findings to those of previous studies.

\citet{Cui12} explored the HMF in simulations with radiative cooling, star formation, and supernova feedback but no AGN feedback\footnote{In the final stages of preparing this paper, \citet{Cui14} posted a paper to the arXiv exploring the effects of AGN feedback on the HMF.  Consistent with our results, they find a shift of $\approx-20\%$ compared to a DM only simulation.}.  For massive haloes, they concluded that the HMF is affected at only the few per cent level.  This is generally consistent with the results of our REF model, which has same physics but with somewhat different subgrid parametrisations.  However, as shown by many previous authors, models that neglect AGN feedback lead to a significant overcooling problem at high masses resulting in groups and clusters with unrealistic properties. 

\cite{Balaguera13} study the effect of baryon physics on the HMF by combining the HMF from dark matter only simulations with the observed trend in baryon fraction with halo mass of local groups and clusters.  They obtain a (negative) difference in the cluster mass function of  10-15\%, depending on which observational data set they use.  This is similar to what we find in our AGN models, which we have shown to reproduce the observations (see Fig.~\ref{fig:mass_fgas} and \cite{LeBrun14}).  An important caveat of this simple method is that by relying on observations this limits the applicability of this method to relatively low redshifts, where the baryon fractions of clusters can be reasonably well measured.  However, even at low redshifts care must be taken to assess the importance of selection effects and mass estimation biases (see \citealt{LeBrun14} for further discussion). 

\cite{Martizzi13b} use and extend the formalism in \cite{Balaguera13} by allowing for the associated expansion/contraction of the dark matter component.  They calibrate their models using a sample of 51 zoom simulations of clusters, as opposed to using observed baryon fractions.  Surprisingly, they find that even with the inclusion of AGN feedback the obtained baryon fraction is very close to universal, in contradiction with recent observations.  The net effect is that they obtain a small (5\%) {\it positive} variation in the HMF for the runs with AGN feedback, in stark contrast with our results. We hypothesise that if their simulations had simultaneously matched the stellar and gas mass fractions of observed groups and clusters, they would have found a similar negative offset in the HMF. 

\cite{Cusworth13} (see also \citealt{Stanek09}) use the Millennium Gas simulations, which include a run with `pre-heating' and cooling (PC) as well as a hybrid simulation (FO, for feedback only) that combines a semi-analytic model of galaxy formation (with AGN feedback) with a non-radiative cosmological hydrodynamical simulation.  Note that both models have been tuned to some degree to match the properties of local groups and clusters.  Both result in a shift in the local HMF of around -15\%, comparable to what we find in our self-consistent AGN models.  Given the relatively large differences in the subgrid implementations of the PC and FO models and our own AGN models, it is plausible that there will be much larger differences in the predictions at higher redshifts.

Thus far we have focused on the high-mass end and the implications for cluster number counts.  At the low-mass end we compare our results with the work of \citet{Sawala12}.  We find good agreement when we use the high-resolution version of the REF simulation in order to be able to resolve smaller haloes.  As already mentioned, our findings suggest an increase in the relative change in mass towards smaller halo masses, although our work suggests a slightly smaller effect due to the fact that the \gimic \ simulation used in \citet{Sawala12} has somewhat more efficient SN feedback (leading to haloes with slightly lower-than-observed stellar mass fractions; see \citealt{McCarthy12}).

\newpage
\section{Summary and Conclusions}
\label{sec:conclusions}

In this paper we have explored the effects of the introduction of important baryon physics associated with galaxy formation on the total mass of haloes, the mass profile up to large radii ($10R_{200}^{\rm{dmonly}}$) and the halo mass function (HMF). In order to isolate the effects of baryon physics, we used several simulations from the OWLS project with identical initial conditions, box size and resolution, starting from only gravitationally interacting particles (DMONLY). On top of that we added gas hydrodynamics, star formation and primordial cooling in the NOSN\_NOZCOOL simulation, an implementation of kinetic SN feedback in the NOZCOOL simulation, introduction of metal-line cooling in the REF simulation and finally the seeding and growth of black holes and AGN feedback in the AGN 8.0 and AGN 8.5 simulations.  We also explored a different prescription for SN feedback in which the mass loading and the wind velocity depend on the local star-forming gas density in the WDENS simulation.  By comparing the results of different simulations, we were able to isolate the importance of different galaxy formation physics on the HMF.

An important aspect of this work was to compare exactly the same set of haloes when different physical processes are introduced in order to isolate the effects of baryons without introducing a bias.  We therefore applied a halo finding and matching algorithm that takes advantage of the fact that the dark matter particles have unique IDs and the simulations all used identical initial conditions.

Using the matched haloes, we compared the relative difference in the total mass of haloes between the DMONLY and the baryonic simulations. Our results span nearly four orders of magnitude in halo mass, $10^{11.5}  < M^{\mathrm{crit}}_{200} / [\hMsun]< 10^{15.2} $.  We found that at the low-mass end, SN feedback produces haloes that are 20\% less massive, at $z=0$, with respect to their DMONLY counterparts, due to the ejection of baryons from the haloes as well as some expansion of the dark matter itself. This difference decreases with increasing halo mass, as the escape velocity gradually becomes too high for the gas to escape, reaching no difference for masses  $M^{\mathrm{crit}}_{200} > 10^{12.5}\hMsun$.The mass range over which SN feedback can alter the HMF can be extended somewhat if the wind velocity increases with local gas density (as in WDENS).  However,
only AGN feedback can produce a substantial alteration of the HMF of galaxy groups and clusters with halo masses up to $M^{\mathrm{crit}}_{200} = 10^{14.8}\hMsun$.

A direct effect of the change in the total mass of the haloes is the modification of the HMF.  Similarly to the total mass variation, we found that supernova feedback is particularly important in shaping the HMF in the mass range $10^{11.5} < M^{\mathrm{crit}}_{200} < 10^{12.5}\hMsun $, with a decrease of $20\%$ with respect to the DM-only simulation. Including only stellar feedback does not produce a significant effect for haloes more massive than $M^{\mathrm{crit}}_{200} = 10^{13}\hMsun$.  In the higher mass range, AGN feedback can induce a similar $20\%$ decrease with respect to the DM-only scenario. 

Baryon physics is able to significantly change the total mass profiles of haloes out to several times the virial radius. This means that also the environment in which the haloes reside has significantly different properties with respect to the simulations with only gravitationally interacting particles. This effect could be very important for gravitational lensing measurements, which are sensitive to the mass profile of haloes \citep[e.g.][]{Mead10,Semboloni11,Killedar12, VanDaalen13}. 

We have provided a set of analytic functions that can be used to correct the masses of DMONLY simulated haloes for the presence of baryons (for several mass definitions).  We have shown that the change in mass of the haloes due to baryon physics does not depend on small changes in the values of the cosmological parameters.
We also used the analytic fitting formulas to correct the Tinker et al.\ universal HMF for the effects of baryons.  In this way we are able to predict the abundance of haloes in different cosmologies.  In particular, we showed that the shift in the HMF is about $20\%$, which has important implication for cluster number counts (e.g., the effect of baryon physics is of the same order as switching the cosmological parameters between WMAP7 and Planck).  To help alleviate this problem, we advocate using only the highest-mass clusters for number counts test, for example $ M^{\rm mean}_{500} >10^{15}\hMsun$, where the effect of baryon physics on the mass of the haloes is far less pronounced.

In conclusion we have shown that the masses of haloes inferred from DM-only simulations are not reliable, and when baryon physics is included this can lead to a difference up to $20\%$ in the mass of the halo and a similar shift in the HMF.  The magnitude of the effect {\it far} exceeds the percentage precision requirement on the HMF \citep{Wu10} for future surveys that aim to constrain the dark energy equation of state, such as XMM-XXL, eROSITA, Planck, DES, Euclid and LSST.  Thus, it is beyond question that baryons must be properly modelled for future precision cosmological studies, as well as for any other theoretical studies that require halo masses to be known with better than $\sim20\%$ accuracy.


\section*{Acknowledgements}  
\label{sec:acknowledgements}
We thank the anonymous referee for insightful comments that helped improve the manuscript. The simulations presented here were run on Stella, the LOFAR Blue Gene/L system in Groningen, on the Cosmology Machine at the Institute for Computational Cosmology in Durham as part of the Virgo Consortium research programme, and on Darwin in Cambridge. This work was sponsored by National Computing Facilities Foundation (NCF) for the use of supercomputer facilities, with financial support from the Netherlands Organization for Scientific Research (NWO). This work was supported by the European Research Council under the European Union's Seventh Framework Programme (FP7/2007-2013) / ERC Grant agreement 278594-GasAroundGalaxies and from the Marie Curie Training Network CosmoComp (PITN-GA-2009- 238356). IGM is supported by an STFC Advanced Fellowship at Liverpool John Moores University. AMCLB acknowledges support from an internally funded PhD studentship at the Astrophysics Research Institute of Liverpool John Moores University. We thank Ming Sun and Yen-Ting Lin for providing their observational data.


\bibliographystyle{mn2e}
\bibliography{paper} 
\bsp

\appendix

\section{Resolution test}
\label{sec:res_test}
\begin{figure*} \begin{center} \begin{tabular}{cc}
\includegraphics[width=0.9\columnwidth ]{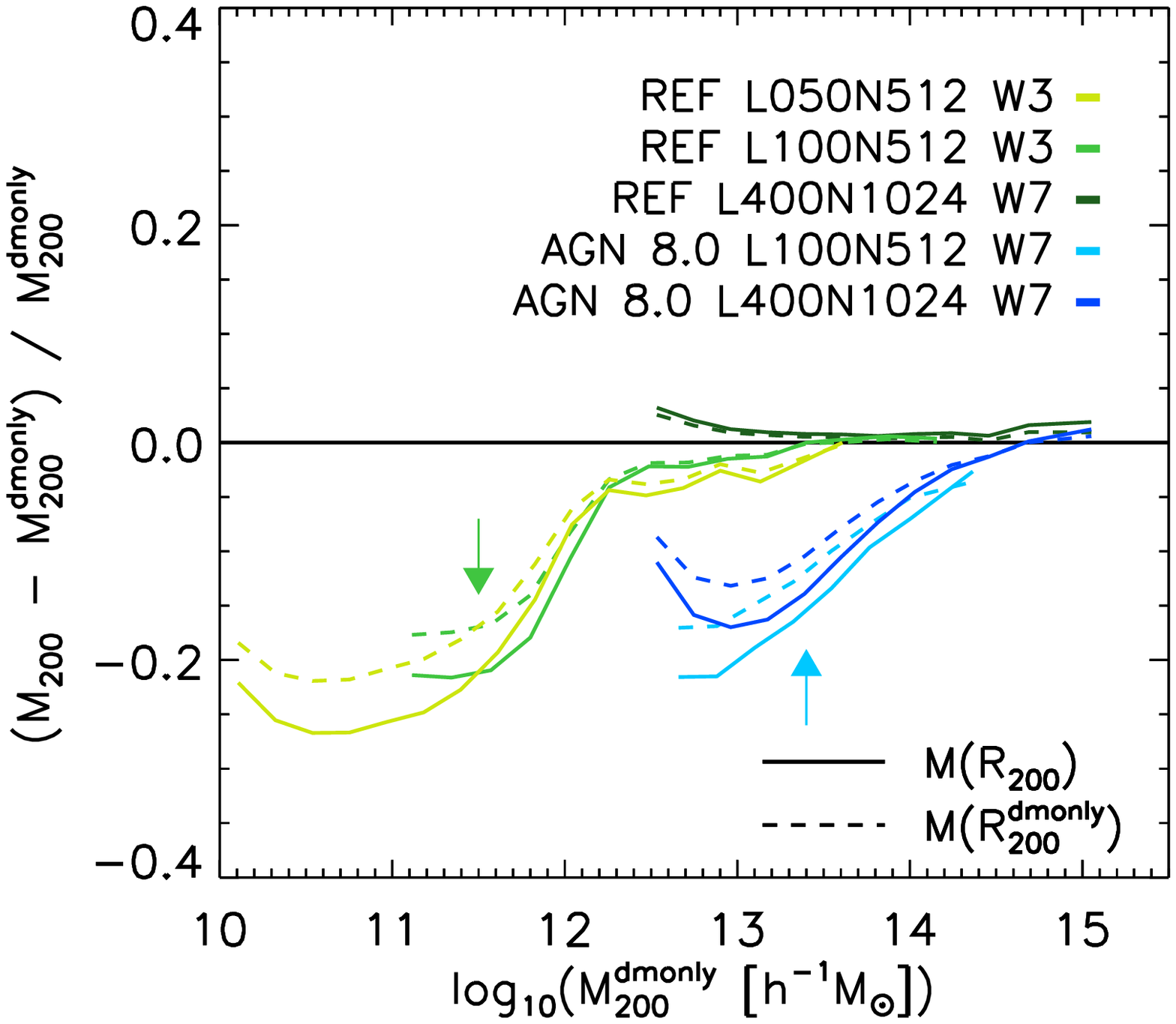} &   {\includegraphics[width=0.9\columnwidth  ]{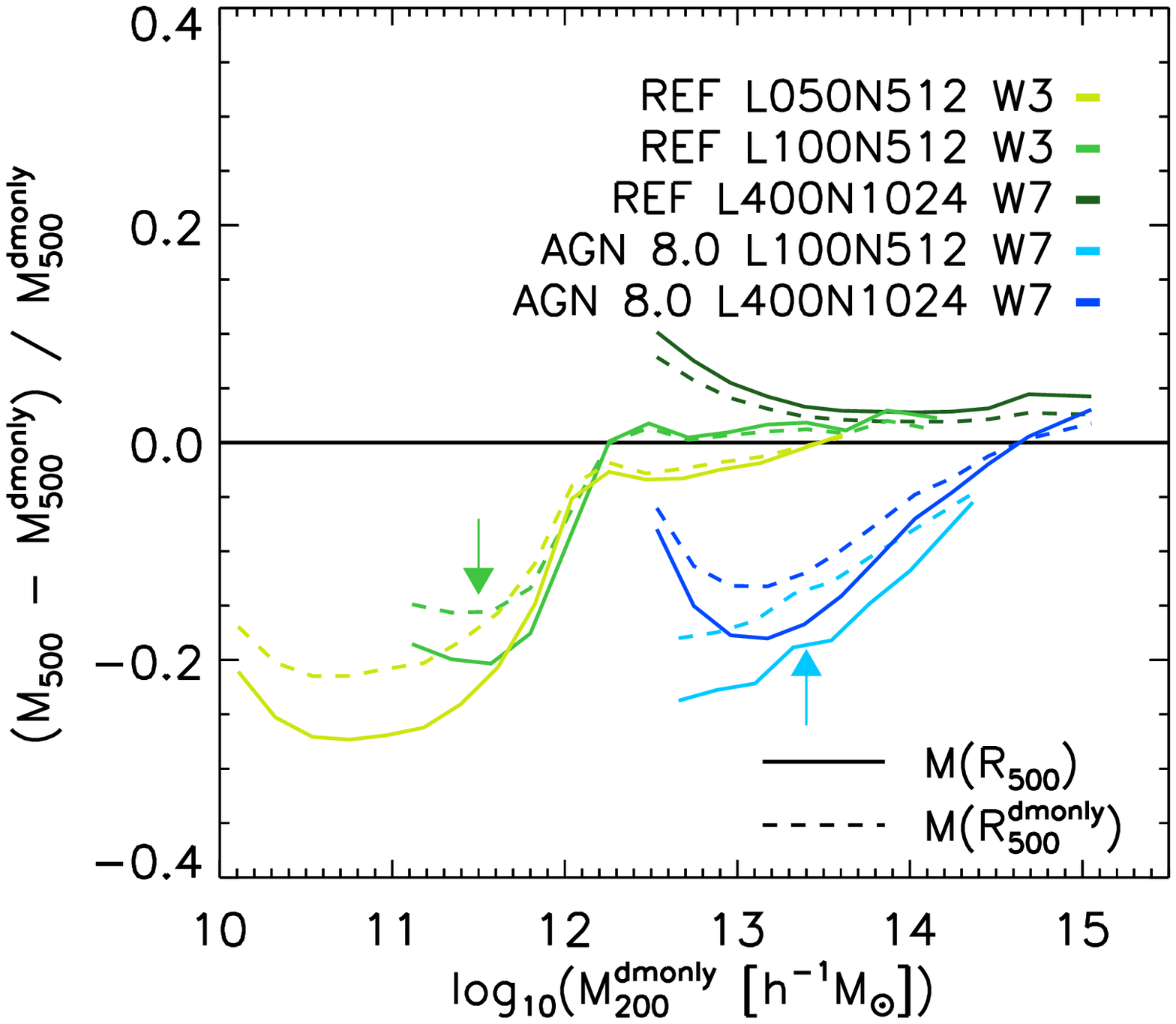}} \\
  {\includegraphics[width=0.9\columnwidth ]{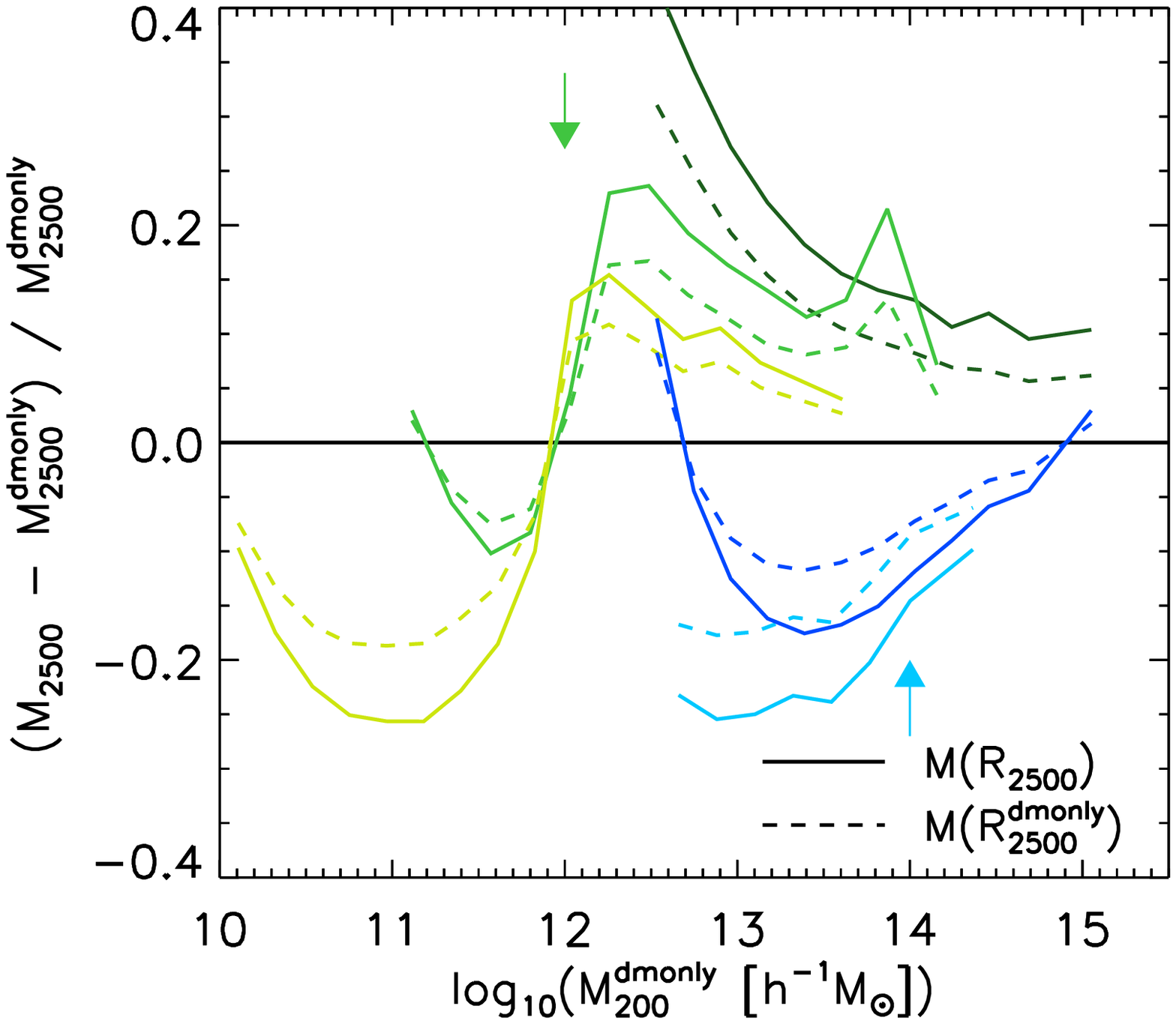}} & {\includegraphics[width=0.9\columnwidth  ]{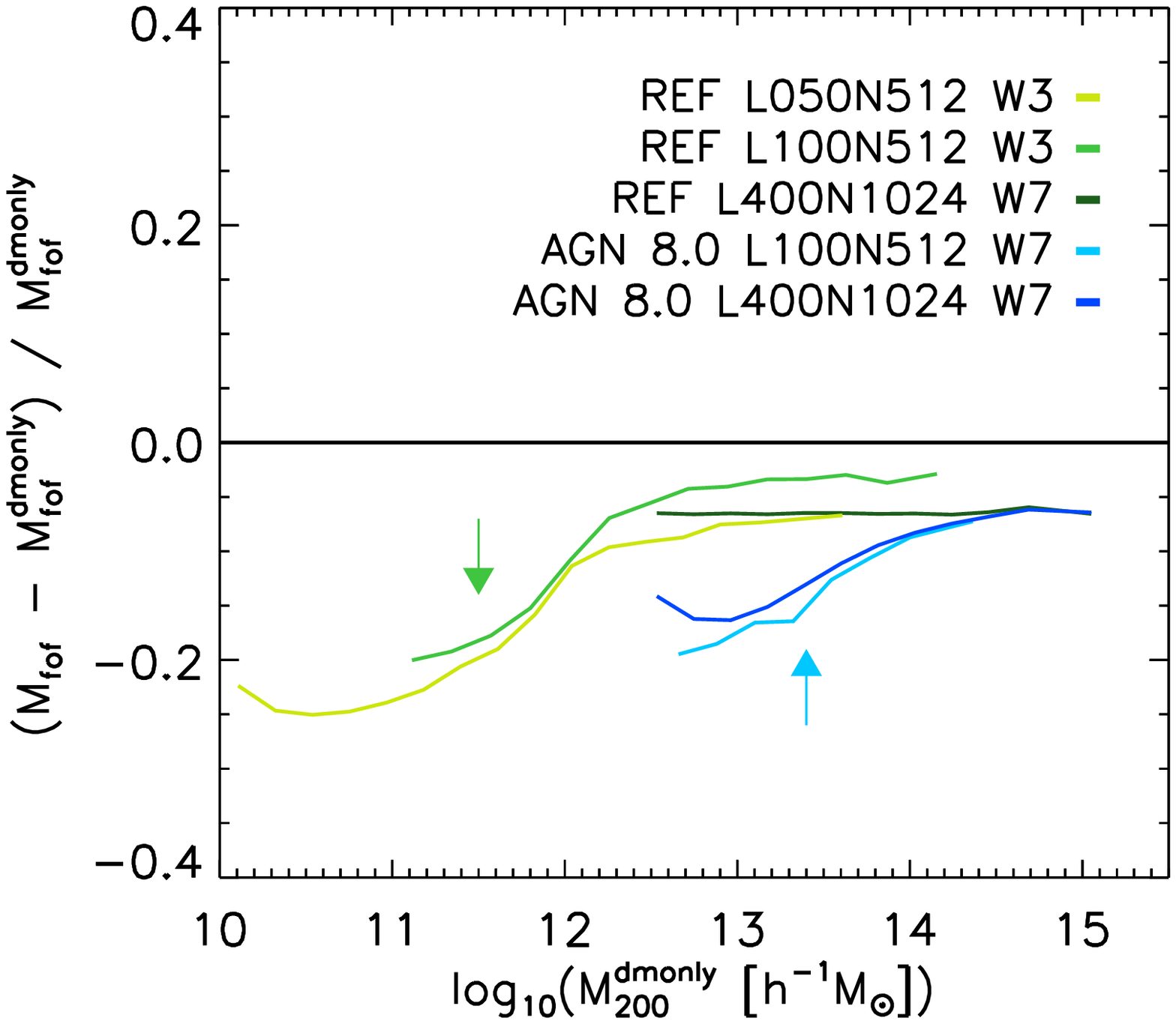}} \\
\end{tabular} \end{center}
\caption{Resolution test for the analysis done in Sec.~\ref{sec:diff}. We show the same panels as in Fig.~\ref{fig:diff200_z0} with the difference that we include only the results from the REF simulations and AGN 8.0, and for every simulation we show the effect of changing the resolution. The arrows show the resolution limits adopted in this work, the arrows that are pointing downward refer to the resolution limit for the simulations done in the 100 $\hMpc$ box, while the upwards pointing arrows show where we  switch to the larger box size.} 
\label{fig:restest}
\end{figure*}

In this section we perform resolution tests for the analysis done in Sec.~\ref{sec:diff}. In Fig.~\ref{fig:restest} we repeat the same analysis performed on the standard resolution (REF green lines) using eight times better mass resolution (REF L050N512) and eight times worse mass resolution with a larger box size (REF L400N1024). We do not have the higher resolution version of the AGN models (blue lines), but we argue that at the low-mass end the behaviour is similar to the REF model since the AGN feedback is not efficient for those low-mass haloes. Instead, we show the effect of a low-resolution version (AGN L400N1024). The high-resolution run and the standard run agree reasonably well for masses  $M_{200}^{\mathrm{dmonly}}  >10^{11.5}\hMsun$, and this is the reason we choose this mass as the lower limit in our analysis. The upturn that is visible at low masses in both simulations is a resolution artefact since it is present at the low-mass end of every simulation but shifted by a factor of eight in mass, i.e. the difference in  mass resolution between the two simulations. The vertical arrows show the chosen resolution limits that approximately correspond to 600 DM particles in the DMONLY simulation with standard resolution.
In the other panels we show resolution tests for the other mass definitions used. We find that $M_{200}^{\mathrm{dmonly}}  =10^{11.5}\hMsun$ is also a good choice for  $M_{500}$ and $M_{\rm fof}$. However, for $M_{2500}$ a better choice for the resolution mass limit is $M_{200}^{\mathrm{dmonly}}  =10^{12}\hMsun$.

\begin{figure*} \begin{center} \begin{tabular}{cc}
\includegraphics[width=0.9\columnwidth ]{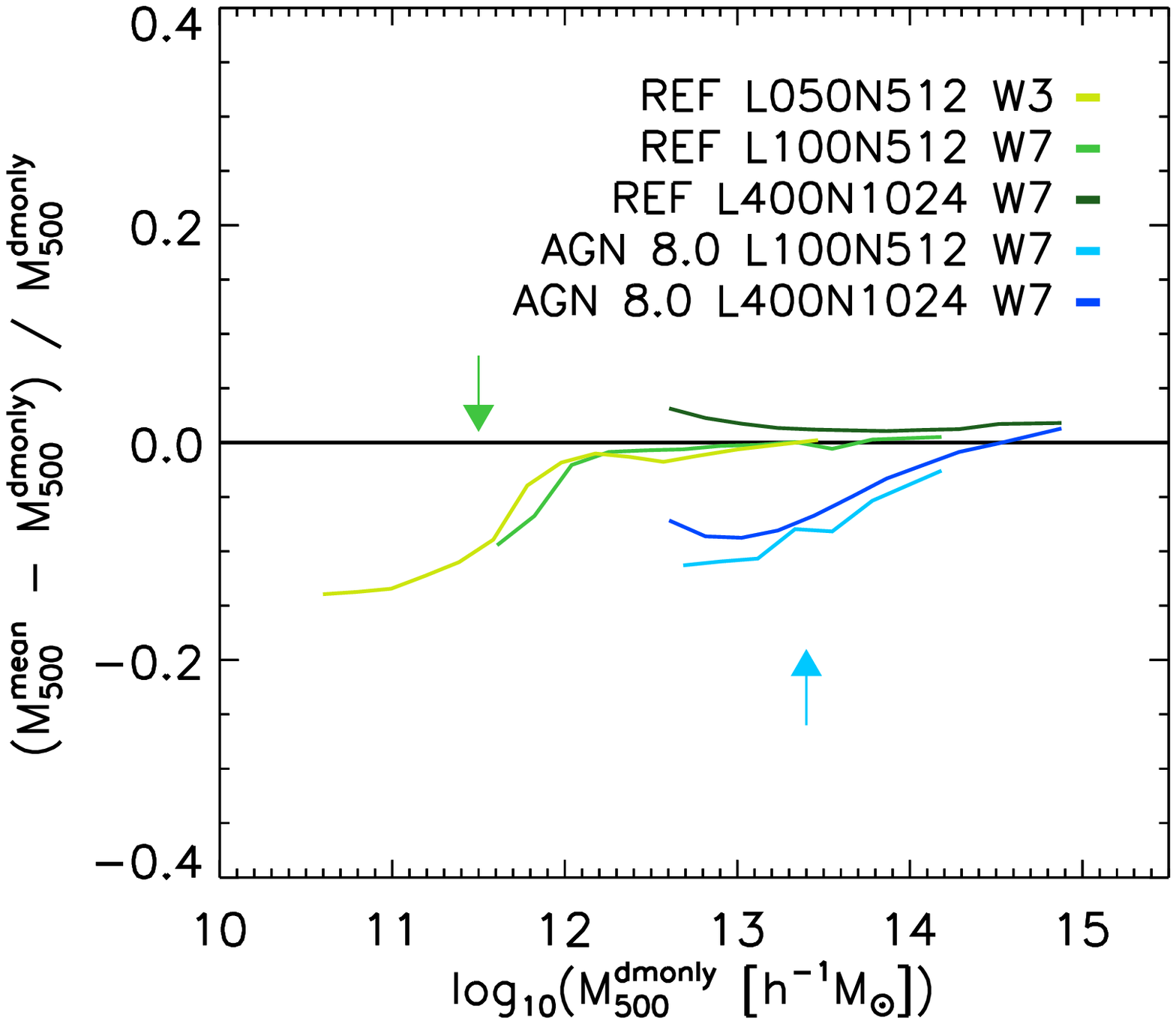} &   {\includegraphics[width=0.9\columnwidth  ]{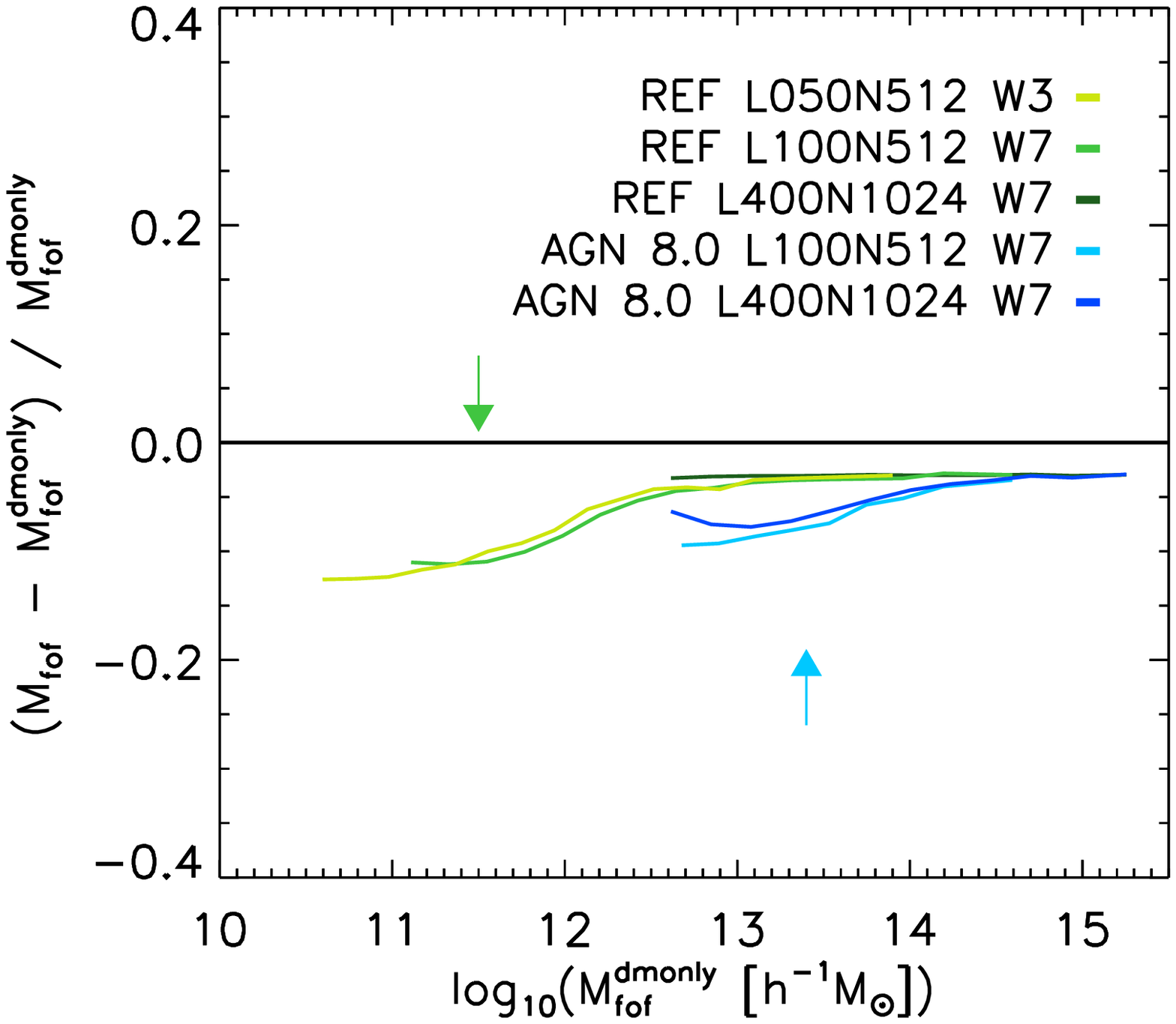}} \\
\end{tabular} \end{center}
\caption{Resolution test for the values used in the fitting functions. We explicitly show the resolution test for $M_{500}^{\rm mean}$ and for $M_{\rm fof}$. The resolution limits are the same as the one used when the haloes are binned in $M_{200}^{\rm crit}$.} 
\label{fig:restest_fit}
\end{figure*}

We also show in Fig.~\ref{fig:restest_fit} the same resolution test for $M_{500}^{\rm mean}$ and for $M_{\rm fof}$ when they are plotted as functions of DMONLY mass with the same mass definition. These relations are used to produce the fitting functions and here we show that the mass resolution limit is the same as in the previous figure where all the quantities are plotted as a function of $M_{200}^{\rm dmonly}$. The same applies to the other mass definitions used in the fitting function.

\label{lastpage}
\end{document}